\documentclass[12pt,pra,aps,amssymb,amsmath,tightenlines]{revtex4}
\usepackage{dsfont}

\newcommand{\half}{\mbox{$\textstyle \frac{1}{2}$}}

\newcommand{\re}{\mbox{$\rm e$}}

\newcommand{\rd}{\mbox{$\rm d$}}

\begin{document}

\title{Information-Based Asset Pricing}

\author{Dorje~C.~Brody${}^*$, Lane~P.~Hughston${}^\dagger$, and
Andrea~Macrina${}^\dagger$}

\affiliation{${}^*$Blackett Laboratory, Imperial College, London
SW7 2BZ, UK \\ ${}^\dagger$Department of Mathematics, King's
College London, The Strand, London WC2R 2LS, UK}


\begin{abstract}
{\bf Abstract}. A new framework for asset price dynamics is
introduced in which the concept of noisy information about future
cash flows is used to derive the corresponding price processes. In
this framework an asset is defined by its cash-flow structure. Each
cash flow is modelled by a random variable that can be expressed as
a function of a collection of independent random variables called
market factors. With each such ``$X$-factor'' we associate a market
information process, the values of which we assume are accessible to
market participants. Each information process consists of a sum of
two terms; one contains true information about the value of the
associated market factor, and the other represents ``noise''. The
noise term is modelled by an independent Brownian bridge that spans
the interval from the present to the time at which the value of the
factor is revealed. The market filtration is assumed to be that
generated by the aggregate of the independent information processes.
The price of an asset is given by the expectation of the discounted
cash flows in the risk-neutral measure, conditional on the
information provided by the market filtration. In the case where the
cash flows are the dividend payments associated with equities, an
explicit model is obtained for the share-price process. Dividend
growth is taken into account by introducing appropriate structure on
the market factors. The prices of options on dividend-paying assets
are derived. Remarkably, the resulting formula for the price of a
European-style call option is of the Black-Scholes-Merton type. We
consider both the case where the rate at which information is
revealed to the market is constant, and the case where the
information rate varies in time. Option pricing formulae are
obtained for both cases. The information-based framework generates a
natural explanation for the origin of stochastic volatility in
financial markets, without the need for specifying on an \textit{ad
hoc} basis the dynamics of the volatility.
\\ \vspace{0.2cm}

\vspace{0.2cm}

\noindent Working paper. Original version: December 5,  2005. This
version: {\today}.

\noindent Email: dorje@imperial.ac.uk, lane.hughston@kcl.ac.uk,
andrea.macrina@kcl.ac.uk
\end{abstract}


\maketitle


\section{Introduction}
\label{sec:1}

In derivative pricing, the starting point is usually the
specification of a model for the price process of the underlying
asset. Such models tend to be of an \textit{ad hoc} nature. For
example, in the Black-Scholes-Merton-Merton theory, the underlying
asset has a geometric Brownian motion as its price process. More
generally, the economy is often modelled by a probability space
equipped with the filtration generated by a multi-dimensional
Brownian motion, and it is assumed that asset prices are adapted to
this filtration. This example is of course the ``standard'' model
within which a great deal of financial engineering has been carried
out. The basic problem with the standard model (and the same applies
to various generalisations thereof) is that the market filtration is
fixed, and no comment is offered on the issue of ``where it comes
from''. In other words, the filtration, which represents the
revelation of information to market participants, is modelled first,
in an \textit{ad hoc} manner, and then it is assumed that the asset
price processes are adapted to it. But no indication is given about
the nature of this ``information'', and it is not obvious, \textit{a
priori}, why the Brownian filtration, for example, should be
regarded as providing information rather than noise.

In a complete market there is a sense in which the Brownian
filtration provides no irrelevant information. That is to say, in a
complete market based on a Brownian filtration the asset price
movements reflect the information content of the filtration.
Nevertheless, the notion that the market filtration should be
``prespecified'' is an unsatisfactory one in financial modelling.
The intuition behind the ``prespecified-filtration'' approach is
that the filtration  represents the unfolding in time of a
succession of random events that ``influence'' the markets, causing
prices to change. For example, bad weather in South America results
in a decrease in the supply of coffee beans and hence an increase in
the price of coffee. Or, say, a spate of bad derivative deals causes
a drop in confidence in banks, and hence a downgrade in earnings
projections, and thus a drop in their prices. The idea is that one
``abstractifies'' these influences in the form of a prespecified
background filtration to which price processes are adapted. What is
unsatisfactory about this is that little structure is given to the
filtration: price movements behave as though they were spontaneous.
In reality, we expect the price-formation process to exhibit more
structure.

It would be out of place in the present context to attempt an
account of the process of price formation. Nevertheless, we can
improve on the ``prespecified'' approach. In that spirit we proceed
as follows. We note that price changes arise from two sources. The
first is that resulting from changes in agent preferences---that is
to say, changes in the pricing kernel. Movements in the pricing
kernel are associated with (a) changes in investor attitudes towards
risk, and (b) changes in investor ``impatience'', the subjective
discounting of future cash flows. Equally important are changes in
price resulting from the revelation of information about the future
cash flows derivable from a given asset. When a market agent decides
to buy or sell an asset, the decision is made in accordance with the
information available to the agent concerning the likely future cash
flows associated with the asset. A change in the information
available to the agent about a future cash flow will typically have
an effect on the price at which they are willing to buy or sell,
even if the agent's preferences remain unchanged. Consider the
situation where one is thinking of purchasing an item at a price
that seems attractive. But then, one reads an article pointing out
an undesirable feature of the product. After reflection, one decides
that the price is too high, given the deficiencies that one is now
aware of. As a result, one decides not to buy, not at that price,
and eventually---because other individuals will have read the same
report---the price drops.

The movement of the price of an asset should, therefore, be regarded
as \emph{an emergent phenomenon}. To put the matter another way, the
price process of an asset should be viewed as the output of (rather
than an input into) the decisions made relating to possible
transactions in the asset, and these decisions should be understood
as being induced primarily by the flow of information to market
participants. Taking into account this observation we propose in
this paper a new framework for asset pricing based on {\it modelling
of the flow of market information}. The information is that
concerning the values of the future cash flows associated with the
given assets. For example, if the asset represents a share in a firm
that will make a single distribution at some agreed date, then there
is a single cash flow. If the asset is a credit-risky discount bond,
then the future cash flow is the payout of the bond at maturity. In
each case, based on the information available relating to the likely
payouts of the given financial instrument, market participants
determine estimates for the value of the right to the impending cash
flows. These estimates lead to the decisions concerning transactions
that trigger movements in the price.

In this paper we present a class of models capturing the essence of
the scenario described above. In building the framework we have
several criteria in mind that we would like to see satisfied. The
first of these is that our model for the flow of market information
should be intuitively appealing, and should allow for a reasonably
sophisticated account of aggregate investor behaviour. At the same
time, the model should be simple enough to allow one to derive
explicit expressions for the asset price processes thus induced, in
a suitably rich range of examples, as well as for various associated
derivative price processes. The framework should also be flexible
enough to allow for the modelling of assets having complex cash-flow
structures. Furthermore, it should be suitable for practical
implementation. The framework should be mathematically sound, and
manifestly arbitrage-free. In what follows we show how our modelling
framework goes a long way towards satisfying these criteria.

The role of information in financial modelling has long been
appreciated, particularly in the theory of market microstructure
(see, e.g., Back \cite{back}, Back and Baruch \cite{back2}, O'Hara
\cite{ohara}, and references cited therein). The present framework
is perhaps most closely related to the line of investigation
represented, e.g., in Cetin \textit{et al}. \cite{cetin}, Duffie and
Lando \cite{duffie2}, Giesecke \cite{giesecke}, Giesecke and
Goldberg \cite{giesecke2}, Guo \textit{et al}. \cite{guo}, and
Jarrow and Protter \cite{jarrow3}. The work in this paper, in
particular, develops that described in Brody \textit{et al}.
\cite{bh1} and Macrina \cite{macrina}, where preliminary accounts of
some of this material appear (see also Rutkowski and Yu
\cite{rutkowski}).

The paper is organised as follows. In Sections~\ref{sec:2},
\ref{sec:x1}, and \ref{sec:x2} we illustrate the framework for
information-based pricing by considering the scenario in which there
is a single random cash flow. An elementary model for market
information is presented, based on the specification of a process
composed of two parts: a ``signal'' component containing true
information about the upcoming cash flow, and an independent
``noise'' component which we model in a specific way. A closed-form
expression for the asset price is obtained in terms of the market
information available at the time the price is being specified. This
result is summarised in Proposition 1. In Section~\ref{sec:3} we
show that the resulting asset price process is driven by a Brownian
motion, an expression for which can be obtained in terms of the
market information process: this construction indicates in explicit
terms the sense in which the price process can be viewed as an
``emergent'' phenomenon. In Section~\ref{sec:4} we show that the
value of a European-style call option, in the case of an asset with
a single cash flow, admits a simple formula analogous to that of the
Black-Scholes-Merton model. In Section~\ref{sec:5} we derive pricing
formulae for the situation when the random variable associated with
the single cash flow has an exponential distribution or, more
generally, a gamma distribution.

The extension of the framework to assets associated with multiple
cash flows is established in Section~\ref{sec:6}. We show that once
the relevant cash flows are decomposed in terms of a collection of
independent market factors, then a closed-form expression for the
asset price associated with a complex cash-flow structure can be
obtained. In Section \ref{sec:x3} we show how the standard geometric
Brownian motion model can be derived in an information-based
setting. This remarkable result motivates the specific choice of
information process given in equation (\ref{eq:2-21}). In Section
\ref{sec:x4} we present a simple model for dividend growth. In
Section \ref{sec:x5} we show that by allowing distinct assets to
share one or more common market factors in the determination of one
or more of their respective cash flows we obtain a natural
correlation structure for the associated asset price processes. This
method for introducing correlation in asset price movements
contrasts with the \textit{ad hoc} approach adopted in most
financial modelling. In Section~\ref{sec:7} we demonstrate that if
two or more market factors affect the future cash flows of an asset,
then the corresponding price process will exhibit unhedgeable
stochastic volatility. This result is noteworthy since even for the
class of relatively simple models considered here it is possible to
identify a candidate for \textit{the origin of stochasticity in
price volatility}, as well as the form it should take, which is
given in Proposition~2.

In the remaining sections of the paper we consider the case where
the rate at which the information concerning the true value of an
impending cash flow is revealed is time dependent. The introduction
of a time-dependent information rate adds additional flexibility to
the modelling framework, and opens the door to the possibility of
calibrating the model to the market prices of options. We consider
the single-factor case first, and obtain a closed-form expression
for the conditional expectation of the cash flow. The result is
stated first in Section~\ref{sec:8} as Proposition~3, and the
derivation is then given in the two sections that follow. In
Section~\ref{sec:9} we introduce a new measure appropriate for the
consideration of a Brownian bridge with a random drift, which is
used in Section~\ref{sec:10} to obtain an expression for the
conditional probability density of the random cash flow. The
consistency of the resulting price process is established in
Section~\ref{sec:11}. We show, in particular, that, for the given
information process, if we re-initialise the model at some specified
future time, the dynamics of the model moving forward from that time
can be represented by a suitably re-initialised information process.
The statement of this result is given in Proposition~4. The
dynamical equation satisfied by the price process is analysed in
Section~\ref{sec:12}, where we demonstrate in Proposition~5 that the
driving process is a Brownian motion, just as in the constant
parameter case. In Section~\ref{sec:14} we derive the price of a
European-style call option in the case for which the information
rate is time dependent.

Our approach is based on the idea that first one models the cash
flows, then the information processes, then the filtration, and
finally the prices. In Section~\ref{sec:15}, we solve the
corresponding ``inverse'' problem. The result is stated in
Proposition~6. Starting from the dynamics of the conditional
distribution of the impending payoff, which is driven by a Brownian
motion adapted to the market filtration, we construct (a) the random
variable that represents the relevant market factor, and (b) an
independent Brownian bridge representing irrelevant information.
These two combine to generate the market filtration. We conclude in
Section~\ref{sec:16} with a general multi-factor extension of the
time-dependent setup, for which the dynamics of the resulting price
processes are given in Propositions 7 and 8.

\section{The Modelling framework}
\label{sec:2}

In asset pricing we require three ingredients: (a) the cash flows,
(b) the investor preferences, and (c) the flow of information to
market participants. Translated into a mathematical language,
these ingredients amount to the following: (a$'$) cash flows are
modelled as random variables; (b$'$) investor preferences are
modelled with the determination of a pricing kernel; and (c$'$)
the market information flow is modelled with the specification of
a filtration. As we have indicated above, asset pricing theory
conventionally attaches more weight to (a) and (b) than to (c). In
this paper we emphasise the importance of ingredient (c).

Our theory will be based on modelling the flow of information
accessible to market participants concerning the future cash flows
associated with a position in a financial asset. We start by
introducing the notation and assumptions employed in this paper.
We model the financial markets with the specification of a
probability space $(\Omega,{\mathcal F}, {\mathbb Q})$ on which a
filtration $\{{\mathcal F}_t\}_{0\leq t < \infty}$ will be
constructed. The probability measure ${\mathbb Q}$ is understood
to be the risk-neutral measure, and the filtration $\{{\mathcal
F}_t\}$ is understood to be the market filtration. All asset-price
processes and other information-providing processes accessible to
market participants will be adapted to $\{{\mathcal F}_t\}$.

Several simplifying assumptions will be made that will enable us to
concentrate our efforts on the problems associated with the flow of
market information. The first of these assumptions is the use of the
risk-neutral measure.  The ``real'' probability measure does not
enter into the present investigation. Expectation in the measure
${\mathbb Q}$ will be denoted by ${\mathbb E}[-]$. Our second
assumption is that we take the default-free system of interest rates
to be deterministic. The absence of arbitrage then implies that the
corresponding system of discount functions $\{P_{tT}\}_{0\le t \leq
T<\infty}$ can be written in the form $P_{tT}= P_{0T}/P_{0t}$ for $t
\leq T$, where $\{P_{0t}\}_{0\le t<\infty}$ is the initial discount
function, which we take to be part of the initial data of the model.
The function $\{P_{0t}\}_{0\leq t< \infty}$ is assumed to be
differentiable and strictly decreasing, and to satisfy $0<P_{0t}\leq
1$ and $\lim_{t\rightarrow\infty}P_{0t}=0$.

We also assume, for simplicity, that cash flows occur at
pre-determined dates. Now clearly for some purposes we would like
to allow for cash flows occurring effectively at random times---in
particular, at stopping times associated with the market
filtration. But in the present exposition we want to avoid the
idea of a ``prespecified'' filtration with respect to which
stopping times are defined. We take the view that the market
filtration is a ``derived'' notion, generated by information about
upcoming cash flows, and by the values of cash flows when they
occur. We shall therefore regard a ``randomly-timed'' cash flow as
being a {\it set} of random cash flows occurring at various times,
and with a joint distribution function that ensures only one of
these flows is non-zero. Hence in our view the ontological status
of a cash flow is that its timing is definite, only the amount is
random---and that cash flows occurring at different times are, by
their nature, different cash flows.

\section{Modelling the cash flows}
\label{sec:x1}

First we consider the case of an asset that provides a single
isolated cash flow occurring at time $T$, represented by a random
variable $D_T$. We assume that $D_T\ge 0$. The value $S_t$ of the
cash flow at any earlier time $t$ in the interval $0\le t< T$ is
then given by the discounted conditional expectation of $D_T$:
\begin{eqnarray}
S_t=P_{tT}{\mathbb E}\left[D_T\vert{\mathcal F}_t\right].
\label{eq:2-1}
\end{eqnarray}
In this way we model the price process $\{S_t\}_{0\le t<T}$ of a
limited-liability asset that pays a single dividend $D_T$ at time
$T$. The construction of the price process here is carried out in
such a way as to guarantee an arbitrage-free market if other
assets are priced by the same method (see Davis \cite{davis} for a
closely related point of view). We shall use the terms ``cash
flow'' and ``dividend'' more or less interchangeably. If a more
specific use of one of these terms is needed, then this will be
evident from the context. We adopt the convention that when the
dividend is paid the asset price goes ``ex-dividend'' immediately.
Hence in the example above we have $\lim_{t\to T}S_{t}=D_T$ and
$S_T=0$.

In the case where the asset pays a sequence of dividends $D_{T_k}$
$(k=1,2,\ldots,n)$ on the dates $T_k$ the price (for $t<T_1$) is
given by
\begin{eqnarray}
S_t=\sum^n_{k=1}P_{tT_k}{\mathbb E}\left[ D_{T_k} \vert{\mathcal
F}_t\right]. \label{eq:2-3}
\end{eqnarray}
More generally, for all $t\ge 0$, and taking into account the
ex-dividend behaviour, we have
\begin{eqnarray}
S_t=\sum^n_{k=1}{\mathds 1}_{\{t<T_k\}}P_{tT_k}{\mathbb E}
\left[D_{T_k}\vert{\mathcal F}_t\right].
\end{eqnarray}

It will be useful  if we adopt the convention that a discount bond
also goes ex-dividend on its maturity date. Thus in the case of a
discount bond we assume that the price of the bond is given, for
dates earlier than the maturity date, by the product of the
principal and the relevant discount factor. But at maturity (when
the principal is paid) the value of the bond drops to zero. In the
case of a coupon bond, there is likewise a downward jump in the
price of the bond at the time a coupon is paid (the value lost may
be captured back in the form of an ``accrued interest'' payment).
In this way we obtain a consistent treatment of the
``ex-dividend'' behaviour of all of the asset price processes
under consideration. With this convention it follows that price
processes are right continuous with left limits.

\section{Modelling the information flow}
\label{sec:x2}

Now we present a simple model for the flow of market information,
following Brody \textit{et al}.~\cite{bh1}. We consider first the
case of a single distribution, occurring at time $T$, and assume
that market participants have only partial information about the
upcoming cash flow $D_{T}$. The information available in the market
about the cash flow is assumed to be contained in a process
$\{\xi_t\}_{0\leq t\leq T}$ defined by:
\begin{eqnarray}
\xi_t=\sigma t D_{T}+\beta_{tT}. \label{eq:2-21}
\end{eqnarray}
We call $\{\xi_t\}$ the \textit{market information process}. This
process is composed of two parts. The term $\sigma t D_T$ contains
the ``true information'' about the dividend, and grows in magnitude
as $t$ increases. The process $\{\beta_{tT}\}_{0\leq t\leq T}$ is a
standard Brownian bridge over the time interval $[0,T]$. Thus
$\{\beta_{tT}\}$ is Gaussian, $\beta_{0T}=0$, $\beta_{TT}=0$, the
random variable $\beta_{tT}$ has mean zero, and the covariance of
$\beta_{sT}$ and $\beta_{tT}$ for $s \leq t$ is $s(T-t)/T$. We
assume that $D_{T}$ and $\{\beta_{tT}\}$ are independent. Thus the
information contained in the bridge process is ``pure noise''.

We assume that the market filtration $\{{\mathcal F}_t\}$ is
generated by the market information process: ${\mathcal F}_t=
\sigma(\{\xi_s\}_{0\leq s\leq t})$. The dividend $D_T$ is therefore
${\mathcal F}_T$-measurable, but is not ${\mathcal F}_t$-measurable
for $t<T$. Thus the value of $D_T$ becomes ``known'' at time $T$,
but not earlier. The bridge process $\{\beta_{tT}\}$ is not adapted
to $\{{\mathcal F}_t\}$ and thus is not directly accessible to
market participants. This reflects the fact that until the dividend
is paid the market participants cannot distinguish the ``true
information" from the ``noise" in the market. The introduction of
the Brownian bridge models the fact that market perceptions, whether
valid or not, play a role in determining asset prices. Initially,
all available information is used to determine the \textit{a priori}
probability distribution for $D_T$. The parameter $\sigma$
represents the rate at which information about the true value of
$D_T$ is revealed as time progresses. If $\sigma$ is low, the value
of $D_T$ is effectively hidden until very near the time of the
dividend payment; whereas if $\sigma$ is high, then the value of the
cash flow is for all practical purposes revealed quickly.

In the example under consideration we have made some simplifying
assumptions concerning the information structure. For instance, we
assume that $\sigma$ is constant. In Section~\ref{sec:8}, however,
we consider a time-dependent information flow rate. We have also
assumed that the random dividend $D_T$ enters directly into the
structure of the information process, and enters linearly. As we
shall indicate later, a more general and natural setup is to let the
information process depend on a random variable $X_T$ which we call
a ``market factor''; then the dividend is regarded as a function of
the market factor. This arrangement has the advantage that it easily
generalises to the situation where a cash flow might depend on
several independent market factors, or indeed where cash flows
associated with different financial instruments have one or more
factors in common.

Given the market information structure described above for a single
cash flow, we proceed to construct the associated price dynamics.
The price process $\{S_t\}$ for a share in the firm paying the
specified dividend is given by formula (\ref{eq:2-1}). It is assumed
that the \textit{a priori} probability distribution of the dividend
$D_T$ is known. This distribution is regarded as part of the initial
data of the problem, which in some cases can be calibrated from
knowledge of the initial price of the asset along with other price
data. The general problem of how the \textit{a priori} distribution
is obtained is an important one---any asset pricing model has to
confront this issue---which we defer for later consideration. The
initial distribution is not to be understood as being ``absolutely''
determined, but rather represents the ``best estimate'' for the
distribution given the data available at that time, in accordance
with what one might call a Bayesian point of view. We note the fact
that the information process $\{\xi_t\}$ is Markovian (see Brody
\textit{et al}. \cite{bh1}, and Rutkowski and Yu \cite{rutkowski}).
Making use of this property together with the fact that $D_T$ is
${\mathcal F}_T$-measurable we deduce that
\begin{eqnarray}
S_t={\mathds 1}_{\{t<T\}}P_{tT}{\mathbb E}\left[D_T|\xi_t\right].
\label{eq:2-7}
\end{eqnarray}
If the random variable $D_T$ that represents the payoff has a
continuous distribution, then the conditional expectation in
(\ref{eq:2-7}) can be expressed in the form
\begin{eqnarray}
{\mathbb E}\left[D_T|\xi_t\right]= \int_0^{\infty} x \pi_t(x)\,
\rd x.
\end{eqnarray}
Here $\pi_t (x)$ is the conditional probability density for the
random variable $D_T$:
\begin{eqnarray}
\pi_t(x)=\frac{\rd}{\rd x}\,{\mathbb Q}(D_T\le x\vert\xi_t).
\end{eqnarray}
We implicitly assume appropriate technical conditions on the
distribution of the dividend that will suffice to ensure the
existence of the expressions under consideration. Also, for
convenience we use a notation appropriate for continuous
distributions, though corresponding results can be inferred for
discrete distributions, or more general distributions, by slightly
modifying the stated assumptions and conclusions.

Bearing in mind these points, we note that the conditional
probability density process for the dividend can be worked out by
use of a form of the Bayes formula:
\begin{eqnarray}
\pi_t(x) = \frac{p(x)\rho(\xi_t|D_T=x)}{\int_0^\infty p(x)
\rho(\xi_t|D_T=x){\rm d}x}. \label{eq:08}
\end{eqnarray}
Here $p(x)$ denotes the {\it a priori} probability density for
$D_T$, which we assume is known as an initial condition, and
$\rho(\xi_t|D_T=x)$ denotes the conditional density for the random
variable $\xi_t$ given that $D_T=x$. Since $\beta_{tT}$ is a
Gaussian random variable with mean zero and variance $t(T-t)/T$, we
deduce that the conditional probability density for $\xi_t$ is
\begin{eqnarray}
\rho(\xi_t|D_T=x) = \sqrt{{\frac{T}{2\pi t(T-t)}}}
\exp\left( -\frac{(\xi_t-\sigma t x)^2T}{2t(T-t)}\right) .
\label{eq:4.13}
\end{eqnarray}
Inserting this expression into the Bayes formula we get
\begin{eqnarray}
\pi_t(x)= \frac{p(x)\exp\left[
\frac{T}{T-t}(\sigma x\xi_t-\tfrac{1}{2}\sigma^2 x^2
t)\right]}{\int^{\infty}_0 p(x)\exp\left[\frac{T}{T-t}(\sigma
x\xi_t-\tfrac{1}{2}\sigma^2 x^2 t)\right]\rd x}.
\end{eqnarray}
We thus obtain the following result for the asset price:

\vspace{0.4cm} \noindent {\bf Proposition~1}. {\it The
information-based price process $\{S_t\}_{0\le t\le T}$ of a
limited-liability asset that pays a single dividend $D_T$ at time
$T$ with a priori distribution
\begin{eqnarray}
{\mathbb Q}(D_T\leq y)=\int_0^y p(x)\, {\rm d}x
\end{eqnarray}
is given by
\begin{eqnarray}
S_t={\mathds 1}_{\{t<T\}} P_{tT}\frac{\int^{\infty}_0 x p(x)
\exp\left[ \frac{T}{T-t}(\sigma x\xi_t-\tfrac{1}{2}\sigma^2 x^2
t)\right]\rd x}{\int^{\infty}_0 p(x)\exp\left[\frac{T}{T-t}(\sigma
x\xi_t-\tfrac{1}{2}\sigma^2 x^2 t)\right]\rd x}, \label{eq:2-11}
\end{eqnarray}
where $\xi_t=\sigma t D_T +\beta_{tT}$ is the market information.}

\section{Asset price dynamics in the case of a single cash flow}
\label{sec:3}

In order to analyse the properties of the price process deduced
above, and to be able to compare it with other models, we need to
work out the dynamics of $\{S_t\}$. One of the advantages of the
model under consideration is that we have an explicit expression for
the price at our disposal. Thus in obtaining the dynamics we need to
find the stochastic differential equation of which $\{S_t\}$ is the
solution. This turns out to be an interesting exercise because it
offers some insights into what we mean by the assertion that market
price dynamics should be regarded as an ``emergent phenomenon''. To
obtain the dynamics associated with the price process $\{S_t\}$ of a
single-dividend paying asset let us write
\begin{eqnarray}\label{eq:3.13a}
D_{tT}={\mathbb E} [D_T\vert\xi_t].
\end{eqnarray}
Evidently, $D_{tT}$ can be expressed in the form
$D_{tT}=D(\xi_t,t)$, where $D(\xi,t)$ is defined by
\begin{eqnarray}
D(\xi,t)=\frac{\int^{\infty}_0 x
p(x)\exp\left[\frac{T}{T-t}(\sigma x\xi-\tfrac{1}{2}\sigma^2 x^2
t)\right]\rd x}{\int^{\infty}_0 p(x)\exp\left[\frac{T}{T-t}(\sigma
x\xi-\tfrac{1}{2}\sigma^2 x^2 t)\right]\rd x}.
\end{eqnarray}
A straightforward calculation making use of the Ito rules shows
that the dynamical equation for $\{D_{tT}\}$ is given by
\begin{eqnarray}
\rd D_{tT} = \frac{\sigma T}{T-t} V_t \left[ \frac{1}{T-t}
\Big(\xi_t - \sigma T D_{tT} \Big) \rd t + \rd \xi_t \right].
\end{eqnarray}
Here $V_t$ is the conditional variance of the dividend:
\begin{eqnarray}
V_t = {\mathbb E}_t\left[\left(D_T-{\mathbb E}_t[D_T]\right)^2
\right] = \int^{\infty}_0 x^2 \pi_t(x)\,\rd x - \left(
\int^{\infty}_0 x \pi_t(x)\,\rd x\right)^2. \label{eq:2.16}
\end{eqnarray}
Therefore, if we define a new process $\{W_t\}_{0\le t<T}$ by
setting
\begin{eqnarray}
W_t=\xi_t-\int^t_0\,\frac{1}{T-s}\Big(\sigma T D_{tT}
-\xi_s\Big)\rd s, \label{eq:2-16}
\end{eqnarray}
we find, after some rearrangement, that
\begin{eqnarray}
\rd D_{tT} =\frac{\sigma T}{T-t}V_t  \rd W_t.
\label{eq:2-17}
\end{eqnarray}
For the dynamics of the asset price we thus have
\begin{eqnarray}
\rd S_t=r_t S_t\rd t+\Gamma_{tT}\rd W_t, \label{eq:2-18}
\end{eqnarray}
where $r_t=-\rd\ln P_{0t}/\rd t$ is the short rate, and the absolute
price volatility $\Gamma_{tT}$ is
\begin{eqnarray}
\Gamma_{tT}=P_{tT}\frac{\sigma T}{T-t}V_t. \label{x3}
\end{eqnarray}

A slightly different way of arriving at this result is as follows.
We start with the conditional probability $\pi_t(x)$. Then, using
the notation above, for its dynamics we obtain
\begin{eqnarray}
\rd\pi_t(x)=\frac{\sigma T}{T-t}(x-D_{tT})\pi_t(x)\,\rd W_t.
\end{eqnarray}
Since according to (\ref{eq:2-7}) the asset price is given by
\begin{eqnarray}
S_t={\mathds 1}_{\{t<T\}}P_{tT}\int_0^{\infty} x \pi_t(x)\,\rd x,
\end{eqnarray}
we can infer the dynamics of $\{S_t\}$ from the dynamics of the
conditional probability $\{\pi_t(x)\}$, once we take into account
formula (\ref{eq:2.16}) for the conditional variance.

As we shall demonstrate later, the process $\{W_t\}$ defined in
(\ref{eq:2-16}) is an $\{{\mathcal F}_t\}$-Brownian motion. Hence
\textit{from the point of view of the market it is the process
$\{W_t\}$ that drives the asset price dynamics}. In this way our
framework resolves the paradoxical point of view usually adopted
in financial modelling in which $\{W_t\}$ is regarded on the one
hand as ``noise'', and yet on the other hand also generates the
market information flow. And thus, instead of hypothesising the
existence of a driving process for the dynamics of the markets, we
are able from the information-based perspective to {\it deduce}
the existence of such a process.

The information-flow parameter $\sigma$ determines the overall
magnitude of the volatility. In fact, $\sigma$ plays a role
analogous to the similarly-labelled parameter in the
Black-Scholes-Merton theory. Thus, we can say that the rate at which
information is revealed in the market determines the magnitude of
the volatility. Everything else being the same, if we increase the
information-flow rate, then the market volatility will increase as
well. According to this point of view, those mechanisms that one
might have thought were destined to make markets more
efficient---e.g., globalisation of the financial markets, reduction
of trade barriers, improved communications, a robust regulatory
environment, and so on---can have the effect of increasing market
volatility, and hence market risk, rather than reducing it.

\section{European-style call options}
\label{sec:4}

Before we turn to the consideration of more general cash flows and
market information structures, let us consider the pricing of a
derivative on an asset for which the price process is governed by
(\ref{eq:2-18}). Specifically, we consider the valuation of a
European call option on such an asset, with strike price $K$, and
exercisable at a fixed date $t$. The option is written on an asset
that pays a single dividend $D_T$ at time $T>t$. The initial value
of the option is
\begin{eqnarray}
C_0=P_{0t}{\mathbb E}\left[(S_t-K)^+\right]. \label{Eopt}
\end{eqnarray}
Inserting the expression for $S_t$ derived in the previous section
into this formula, we obtain
\begin{eqnarray}
C_0=P_{0t}\,{\mathbb E}\left[ \left( P_{tT}\int^{\infty}_0
x\,\pi_t(x)\rd x-K \right)^+\right]. \label{eq:3-2}
\end{eqnarray}
For convenience we write the conditional probability $\pi_t(x)$ in
the form
\begin{eqnarray}
\pi_t(x)=\frac{p_t(x)}{\int^{\infty}_0\,p_t(x)\rd x},
\label{eq:3-3}
\end{eqnarray}
where the ``unnormalised'' density $p_t(x)$ is defined by
\begin{eqnarray}
p_t(x)=p(x)\exp\left[\frac{T}{T-t}\left(\sigma x \xi_t
-\tfrac{1}{2}\sigma^2 x^2 t\right)\right]. \label{eq:3-4}
\end{eqnarray}
Substituting (\ref{eq:3-4}) into (\ref{eq:3-2}) we find that the
value of the option is
\begin{eqnarray}\label{eq:3-3a}
C_0=P_{0t}{\mathbb E}\left[ \frac{1}{\Phi_t}
\left(\int^{\infty}_0\left(P_{tT}x-K\right)p_t(x)\rd
x\right)^+\right],
\end{eqnarray}
where
\begin{eqnarray}
\Phi_t=\int^{\infty}_0 p_t(x)\rd x.
\end{eqnarray}
The random variable $1/\Phi_t$ can be used to introduce a measure
${\mathbb B}$ on $(\Omega,{\mathcal F}_t)$, which we call the
``bridge measure''. The option price can then be written:
\begin{eqnarray}
C_0=P_{0t}{\mathbb E}^{{\mathbb B}} \left[\left( \int^{\infty}_0
\left(P_{tT}x-K\right)p_t(x)\rd x\right)^+\right]. \label{eq:2.29}
\end{eqnarray}
The special feature of the bridge measure, as we establish in
Section~\ref{sec:9} in a more general context, is that the random
variable $\xi_t$ is Gaussian under ${\mathbb B}$. In particular,
under ${\mathbb B}$ we find that $\{\xi_t\}$ has mean $0$ and
variance $t(T-t)/T$. Since $p_t(x)$ can be expressed as a function
of $\xi_t$, when we calculate the expectation in (\ref{eq:2.29}) we
obtain a tractable formula for $C_0$.

To determine the value of the option we define a constant $\xi^*$
(the critical value) by the following condition:
\begin{eqnarray}
\int^{\infty}_0\left(P_{tT}x-K\right)p(x)\exp\left[\frac{T}{T-t}
\left(\sigma x \xi^*-\tfrac{1}{2}\sigma^2 x^2 t\right)\right]\rd
x=0.
\end{eqnarray}
Then the expectation in (\ref{eq:2.29}) can be performed and we find
that the option price is
\begin{eqnarray}\label{eq:ec1}
C_0=P_{0T}\int^{\infty}_0 x\,p(x)\,N\Big(-z^*+\sigma
x\sqrt{\tau}\Big)\rd x-P_{0t}K\int^{\infty}_0 p(x) \,N
\Big(-z^*+\sigma x\sqrt{\tau}\Big)\rd x,
\end{eqnarray}
where $N(x)$ is the standard normal distribution function, and
\begin{eqnarray}
\tau=\frac{tT}{T-t},\qquad z^*=\xi^*\sqrt{\frac{T}{t(T-t)}}\, .
\end{eqnarray}
We see that a tractable expression of the
Black-Scholes-Merton-Merton type is obtained. The option pricing
problem, even for general $p(x)$, reduces to an elementary numerical
problem. It is interesting to note that although the probability
distribution for the price $S_t$ is not of a ``standard'' type,
nevertheless the option valuation problem remains a solvable one.

\section{Examples of specific dividend structures}
\label{sec:5}

In this section we consider the dynamics of assets with various
dividend structures. First we look at a simple asset for which the
cash flow is exponentially distributed. The \textit{a priori}
probability density for $D_T$ is thus of the form
\begin{eqnarray}
p(x)=\frac{1}{\delta}\,\exp\left(-x/\delta\right),
\end{eqnarray}
where $\delta$ is a constant. We can regard the idea of an
exponentially distributed payout as a model for the situation where
little is known about the probability distribution of the dividend,
apart from its mean. Then from formula (\ref{eq:2-11}) we find that
the asset price is:
\begin{eqnarray}
S_t = {\mathds 1}_{\{t<T\}} P_{tT}\frac{\int^{\infty}_0
x\exp(-x/\delta)\exp\left[\frac{T}{T-t}(\sigma x\xi_t-
\tfrac{1}{2}\sigma^2 x^2 t)\right]\rd
x}{\int^{\infty}_0\exp(-x/\delta)\exp\left[\frac{T}{T-t}(\sigma
x\xi_t-\tfrac{1}{2}\sigma^2 x^2 t)\right]\rd x}.
\end{eqnarray}
We note that $S_0=P_{0T}\delta$, so we can calibrate $\delta$ by
use of the initial price. The integrals in the numerator and
denominator in the expression above can be worked out explicitly.
Hence, we obtain a closed-form expression for the price in the
case of a simple asset with an exponentially-distributed cash
flow:
\begin{eqnarray}
S_t={\mathds 1}_{\{t<T\}} P_{tT}\left[\frac{\exp
\left(-\tfrac{1}{2}B_t^2/A_t\right)} {\sqrt{2\pi A_t}\
N(B_t/\sqrt{A_t})}+\frac{B_t}{A_t}\right],
\end{eqnarray}
where $A_t=\sigma^2tT/(T-t)$ and $B_t= \sigma T\xi_t/
(T-t)-\delta^{-1}$.

Next we consider the case of an asset for which the single
dividend paid at $T$ is gamma-distributed. More specifically, we
assume the probability density is of the form
\begin{eqnarray}
p(x)=\frac{\delta^n}{(n-1)!}\,x^{n-1}\exp(-\delta x),
\end{eqnarray}
where $\delta$ is a positive real number and $n$ is a positive
integer. This choice for the probability density also leads to a
closed-form expression for the share price. We find that
\begin{eqnarray}
S_t={\mathds 1}_{\{t<T\}}P_{tT}\frac{\sum\limits^n_{k=0} {n\choose
k} A_t^{\frac{1}{2}k-n}B_t^{n-k}F_k(-B_t/\sqrt{A_t})}
{\sum\limits^{n-1}_{k=0}{n-1 \choose k}
A_t^{\frac{1}{2}k-n+1}B_t^{n-k-1} F_k(-B_t/\sqrt{A_t})},
\end{eqnarray}
where $A_t$ and $B_t$ are as above, and
\begin{eqnarray}
F_k(x)=\int^{\infty}_x\,z^k\exp\left(-\tfrac{1}{2}z^2\right)dz.
\end{eqnarray}
A recursion formula can be worked out for the function $F_k(x)$.
This is given by
\begin{eqnarray}
(k+1)F_k(x)=F_{k+2}(x)-x^{k+1}\exp\left(-\tfrac{1}{2}x^2\right),
\end{eqnarray}
from which it follows that $F_0(x) = \sqrt{2\pi} N(-x)$, $F_1(x) =
\re^{-\frac{1}{2}x^2}$, $F_2(x) = x \re^{-\frac{1}{2}x^2} +
\sqrt{2\pi} N(-x)$, $F_3(x) = (x^2+2) \re^{-\frac{1}{2}x^2}$, and
so on. In general, the polynomial parts of $\{F_k(x)
\}_{k=0,1,2,\ldots}$ are related to the Legendre polynomials.

\section{Market factors and multiple cash flows}
\label{sec:6}

In this section we proceed to consider the more general situation
where the asset pays multiple dividends. This will allow us to
consider a wider range of financial instruments. Let us write
$D_{T_k}$ $(k=1,\ldots,n)$ for a set of random dividends paid at
the pre-designated dates $T_k$ $(k=1,\ldots,n)$. Possession of the
asset at time $t$ entitles the bearer to the cash flows occurring
at times $T_k>t$. For simplicity we assume $n$ is finite. For each
value of $k$ we introduce a set of independent random variables
$X^{\alpha}_{T_k}$ $(\alpha=1,\ldots,m_k)$, which we call market
factors or $X$-factors. For each value of $\alpha$ we assume that
the market factor $X^{\alpha}_{T_k}$ is ${\mathcal
F}_{T_k}$-measurable, where $\{{\mathcal F}_t\}$ is the market
filtration.

For each value of $k$, the market factors $\{X^{\alpha}_{T_j}
\}_{j\le k}$ represent the independent elements that determine the
cash flow occurring at time $T_k$. Thus for each value of $k$ the
cash flow $D_{T_k}$ is assumed to have the following structure:
\begin{eqnarray}
D_{T_k}=\Delta_{T_k}(X^{\alpha}_{T_1},
X^{\alpha}_{T_2},...,X^{\alpha}_{T_k}),
\end{eqnarray}
where $\Delta_{T_k}(X^{\alpha}_{T_1},
X^{\alpha}_{T_2},...,X^{\alpha}_{T_k})$ is a function of
$\sum_{j=1}^k m_j$ variables. For each cash flow it is, so to
speak, the job of the financial analyst (or actuary) to determine
the relevant independent market factors, and the form of the
cash-flow function $\Delta_{T_k}$ for each cash flow. With each
market factor $X^{\alpha}_{T_k}$ we associate an information
process $\{\xi^{\alpha}_{tT_k}\}_{0\leq t\leq T_k}$ of the form
\begin{eqnarray}
\xi^{\alpha}_{tT_k}=\sigma^{\alpha}_{T_k}X^{\alpha}_{T_k}t+
\beta^{\alpha}_{tT_k}.
\end{eqnarray}
Here $\sigma^{\alpha}_{T_k}$ is a parameter, and
$\{\beta^{\alpha}_{tT_k}\}$ is a standard Brownian bridge over the
interval $[0,T_k]$. We assume that the $X$-factors and the Brownian
bridge processes are all independent. The parameter
$\sigma^{\alpha}_{T_k}$ determines the rate at which the market
factor $X^{\alpha}_{T_k}$ is revealed. The Brownian bridge
represents the associated noise. We assume that the market
filtration $\{{\mathcal F}_t\}$ is generated by the totality of the
independent information processes $\{\xi^{\alpha}_{tT_k}\}_{0\leq
t\leq T_k}$ for $k=1,2,\ldots,n$ and $\alpha=1,2,\ldots,m_k$. Hence,
the price of the asset is given by
\begin{eqnarray}
S_t=\sum^n_{k=1}{\mathds 1}_{\{t<T_k\} }P_{tT_k}{\mathbb E}_t
\left[D_{T_k}\right]. \label{eq:42-1}
\end{eqnarray}

\section{Geometric Brownian motion model}
\label{sec:x3}

The simplest application of the $X$-factor technique arises in the
case of geometric Brownian motion models. We consider a
limited-liability company that makes a single cash distribution
$S_T$ at time $T$. We assume that $S_T$ has a log-normal
distribution under ${\mathbb Q}$, and can be written in the form
\begin{eqnarray}
S_T=S_0 \exp\left(rT+\nu\sqrt{T}X_T -\half \nu^2T\right),
\label{eq:zz1}
\end{eqnarray}
where the market factor $X_T$ is normally distributed with mean zero
and variance one, and $r>0$ and $\nu>0$ are constants. The
information process $\{\xi_t\}$ is taken to be of the form
\begin{eqnarray}
\xi_t=\sigma t X_T + \beta_{tT}, \label{eq:zz2}
\end{eqnarray}
where the Brownian bridge $\{\beta_{tT}\}$ is independent of $X_T$,
and where the information flow rate is of the special form
\begin{eqnarray}
\sigma = \frac{1}{\sqrt{T}}.  \label{eq:zz3}
\end{eqnarray}
By use of the Bayes formula we find that the conditional probability
density is of the Gaussian form:
\begin{eqnarray}
\pi_t(x) = \sqrt{\frac{T}{2\pi(T-t)}}\,\exp\left( -\frac{1}{2(T-t)}
\left( \sqrt{T}x-\xi_t\right)^2\right), \label{eq:zz4}
\end{eqnarray}
and has the following dynamics
\begin{eqnarray}
\rd \pi_t(x) = \frac{1}{T-t}\, \left(\sqrt{T}x-\xi_t\right) \pi_t(x)
\rd \xi_t.  \label{eq:zz5}
\end{eqnarray}
A short calculation then shows that the value of the asset at time
$t<T$ is given by
\begin{eqnarray}
S_t &=& \re^{-r(T-t)}{\mathbb E}_t[S_T] \nonumber \\ &=&
\re^{-r(T-t)} \int_{-\infty}^\infty S_0
\re^{rT+\nu\sqrt{T}x-\frac{1}{2}\,\nu^2T} \pi_t(x) \rd x \nonumber \\
&=& S_0 \exp\left(rt+\nu \xi_t-\half\nu^2t \right) . \label{eq:zz6}
\end{eqnarray}
The surprising fact in this example is that $\{\xi_t\}$ itself turns
out to be the innovation process. Indeed, it is not too difficult to
verify that $\{\xi_t\}$ is an $\{{\mathcal F}_t\}$-Brownian motion.
Hence, setting $W_t=\xi_t$ for $0\leq t< T$ we obtain the standard
geometric Brownian motion model:
\begin{eqnarray}
S_t=S_0 \exp\left(rt+\nu W_t-\half\nu^2t \right).  \label{eq:zz7}
\end{eqnarray}
We see therefore that starting with an information process of the
form (\ref{eq:zz2}) we are able to recover the familiar asset price
dynamics given by (\ref{eq:zz7}).

An important point to note here is that the Brownian bridge process
$\{\beta_{tT}\}$ appears quite naturally in this context. In fact,
if we start with (\ref{eq:zz7}) then we can make use of the
following orthogonal decomposition of the Brownian motion (see,
e.g., Yor~\cite{yor}):
\begin{eqnarray}
W_t = \frac{t}{T}\,W_T + \left( W_t-\frac{t}{T}\,W_T\right).
\label{eq:46}
\end{eqnarray}
The second term in the right, independent of the first term on the
right, is a standard representation for a Brownian bridge process:
\begin{eqnarray}
\beta_{tT} = W_t - \frac{t}{T}\,W_T.
\end{eqnarray}
Thus by writing $X_T= W_T/ \sqrt{T}$ and $\sigma=1/\sqrt{T}$ we find
that the right side of (\ref{eq:46}) is indeed the market
information. In other words, formulated in the information-based
framework, the standard Black-Scholes-Merton theory can be expressed
in terms of a normally distributed $X$-factor and an independent
Brownian bridge noise process.

\section{Dividend growth}
\label{sec:x4}

As an elementary example of a multi-dividend structure, we shall
look at a simple growth model for dividends in the equity markets.
We consider an asset that pays a sequence of dividends $D_{T_k}$,
where each dividend date has an associated $X$-factor. Let
$\{X_{T_k}\}_{k=1,\ldots,n}$ be a set of independent,
identically-distributed $X$-factors, each with mean $1+g$. The
dividend structure is assumed to be of the form
\begin{eqnarray}
D_{T_k}=D_0\prod^k_{j=1}X_{T_j},
\end{eqnarray}
where $D_0$ is a constant. The parameter $g$ can be interpreted as
the dividend growth factor, and $D_0$ can be understood as
representing the most recent dividend before time zero. For the
price of the asset we have:
\begin{eqnarray}
S_t=D_0\sum^n_{k=1}{\mathds 1}_{\{t<T_k\}}P_{tT_k}{\mathbb E}_t
\left[\prod^k_{j=1}X_{T_j}\right].
\end{eqnarray}
Since the $X$-factors are independent, the conditional expectation
of the product appearing in this expression factorises into a
product of conditional expectations, and each such conditional
expectation can be written in the form of an expression of the
type we have already considered. As a consequence we are led to a
tractable family of dividend growth models.

\section{Assets with common factors}
\label{sec:x5}

The multiple-dividend asset pricing model introduced in
Section~\ref{sec:6} can be extended in a very natural way to the
situation where two or more assets are being priced. In this case
we consider a collection of $N$ assets with price processes
$\{S^{(i)}_t\}_{i=1,2,\ldots,N}$. With asset number $(i)$ we
associate the cash flows $\{D^{(i)}_{T_k}\}$ paid at the dates
$\{T_k\}_{k=1,2,\ldots,n}$. We note that the dates
$\{T_k\}_{k=1,2,\ldots,n}$ are not tied to any specific asset, but
rather represent the totality of possible cash-flow dates of any
of the given assets. If a particular asset has no cash flow on one
of the dates, then it is assigned a zero cash-flow for that date.
From this point, the theory proceeds exactly as in the single
asset case. That is to say, with each value of $k$ we associate a
set of $X$-factors $X^{\alpha}_{T_k}$ $(\alpha=1,2,\ldots,m_k)$,
and a system of market information processes
$\{\xi^{\alpha}_{tT_k}\}$. The $X$-factors and the information
processes are not tied to any particular asset. The cash flow
$D^{(i)}_{T_k}$ occurring at time $T_k$ for asset number $(i)$ is
given by a cash flow function of the form
\begin{eqnarray}
D^{(i)}_{T_k}=\Delta^{(i)}_{T_k}(X^{\alpha}_{T_1},
X^{\alpha}_{T_2},...,X^{\alpha}_{T_k}).
\end{eqnarray}
In other words, for each asset each cash flow can depend on all of
the $X$-factors that have been ``activated'' at that point. Thus
for the general multi-asset model we have the following price
process system:
\begin{eqnarray}
S^{(i)}_t=\sum^n_{k=1}{\mathds 1}_{\{t<T_k\}} P_{tT_k} {\mathbb
E}_t \left[D^{(i)}_{T_k}\right].
\end{eqnarray}

It is possible in general for two or more assets to ``share'' an
$X$-factor in association with one or more of the cash flows of
each of the assets. This in turn implies that the various assets
will have at least one Brownian motion in common in the dynamics
of their price processes. We thus obtain a natural model for the
correlation structures in the prices of these assets. The
intuition is that as new information comes in (whether ``true'' or
``bogus'') there will be several different assets all affected  by
the news, and as a consequence there will be a correlated movement
in their prices.

\section{Origin of unhedgeable stochastic volatility}
\label{sec:7}

Based on the general model introduced in the previous sections, we
are now in a position to make an observation concerning the nature
of stochastic volatility in the equity markets. In particular, we
shall show how a natural framework for \textit{stochastic
volatility} arises in the information-based framework. This is
achieved without the need for any \textit{ad hoc} assumptions
concerning the dynamics of the stochastic volatility. In fact, a
very specific dynamical model for stochastic volatility is
obtained---thus leading to a possible means by which the theory
proposed here might be tested.

We shall work out the volatility associated with the dynamics of
the asset price process $\{S_t\}$ given by (\ref{eq:42-1}). The
result is given in Proposition~2 below. First, as an example, we
consider the dynamics of an asset that pays a single dividend
$D_T$ at $T$. We assume that the dividend depends on the market
factors $\{X^{\alpha}_T \}_{\alpha=1,\ldots,m}$. For $t<T$ we then
have:
\begin{eqnarray}
S_t&=&P_{tT}{\mathbb E}^{{\mathbb Q}}
\left[\left.\Delta_{T}\left(X^1_{T},\ldots,X^m_{T}\right)
\right|\xi^1_{tT},\ldots,\xi^m_{tT}\right] \nonumber \\
&=& P_{tT}\int\cdots\int\Delta_{T}(x^1,\ldots,x^m)\, \pi^1_{tT}
(x_1)\cdots\pi^{m}_{tT}(x_{m})\,\rd x_1\cdots \rd x_{m}.
\end{eqnarray}
Here the various conditional probability density functions
$\pi^{\alpha}_{tT}(x)$ for $\alpha=1,\ldots,m$ are
\begin{eqnarray}
\pi^{\alpha}_{tT}(x)=\frac{p^{\alpha}(x)\exp\left[\frac{T}{T-t}
\left(\sigma^{\alpha}\,x\,\xi^{\alpha}_{tT}-\tfrac{1}{2}
(\sigma^{\alpha})^2\,x^2 t\right)\right]}
{\int^{\infty}_{0}\,p^{\alpha}(x)
\exp\left[\frac{T}{T-t}\left(\sigma^{\alpha}\,x\,
\xi^{\alpha}_{tT}-\tfrac{1}{2}(\sigma^{\alpha})^2\,x^2
t\right)\right]\rd x},
\end{eqnarray}
where $p^\alpha(x)$ denotes the \textit{a priori} probability
density function for the factor $X_T^\alpha$. The drift of
$\{S_t\}_{0\leq t<T}$ is given by the short rate. This is because
${\mathbb Q}$ is the risk-neutral measure, and no dividend is paid
before $T$. Thus, we are left with the problem of determining the
volatility of $\{S_t\}$. We find that for $t<T$ the dynamical
equation of $\{S_t\}$ assumes the form:
\begin{eqnarray}
\rd S_t=r_t S_t\rd t+\sum^m_{\alpha=1}\Gamma^{\alpha}_{tT}\rd
W^{\alpha}_t. \label{eq:50}
\end{eqnarray}
Here the volatility term associated with factor number $\alpha$ is
given by
\begin{eqnarray}
\Gamma^{\alpha}_{tT}=\sigma^{\alpha} \frac{T}{T-t}
P_{tT}\,\textrm{Cov}\left[\left.\Delta_{T}\left(
X^{1}_{T},\ldots,X^{m}_{T}\right),X^{\alpha}_{T} \right|{\mathcal
F}_t\right], \label{vol}
\end{eqnarray}
and $\{W_t^\alpha\}$ denotes the Brownian motion associated with
the information process $\{\xi_t^\alpha\}$, as defined in
(\ref{eq:2-16}). The absolute volatility of $\{S_t\}$ is of the
form
\begin{eqnarray}
\Gamma_t=\left(\textstyle{\sum\limits_{\alpha=1}^m}
\left(\Gamma^{\alpha}_{tT} \right)^2\right)^{1/2}.
\end{eqnarray}
For the dynamics of a multi-factor single-dividend paying asset we
can thus write $\rd S_t=r_t S_t\rd t+\Gamma_t \rd Z_t$, where the
$\{{\mathcal F}_t\}$-Brownian motion $\{Z_t\}$ that drives the
asset-price process is
\begin{eqnarray}
Z_t=\int_0^t\frac{1}{\Gamma_s}\sum^m_{\alpha=1}
\Gamma^{\alpha}_{sT}\,\rd W^{\alpha}_{s}.
\end{eqnarray}
The point to note here is that in the case of a multi-factor model
we obtain an unhedgeable stochastic volatility. That is to say,
although the asset price is in effect driven by a single Brownian
motion, its volatility in general depends on a multiplicity of
Brownian motions. This means that in general an option position
cannot be hedged with a position in the underlying asset. The
components of the volatility vector are given by the covariances of
the cash flow and the independent market factors. Unhedgeable
stochastic volatility thus emerges from the multiplicity of
uncertain elements in the market that affect the value of the future
cash flow. As a consequence we see that \emph{in this framework we
obtain a natural explanation for the origin of stochastic
volatility}.

This result can be contrasted with, say, the Heston model
\cite{heston}, which despite its popularity suffers from the fact
that it is \textit{ad hoc} in nature. Much the same can be said for
the various generalisations of the Heston model used in commercial
applications. The approach to stochastic volatility proposed in the
present paper is thus of a new character. Expression (\ref{eq:50})
generalises to the case for which the asset pays a set of dividends
$D_{T_k}$ $(k=1, \ldots,n)$, and for each $k$ the dividend depends
on the $X$-factors $\{\{
X^\alpha_{T_j}\}^{\alpha=1,\ldots,m_j}_{j=1, \ldots,k}\}$. The
result can be summarised as follows.

\vspace{0.4cm} \noindent {\bf Proposition~2}. {\it The price
process of a multi-dividend asset has the following dynamics:
\begin{eqnarray}
\rd S_t &=& r_t\,S_t\,\rd t+\sum^n_{k=1}\sum^{m_k}_{\alpha=1}
{\mathds 1}_{\{t<T_k\}} \frac{\sigma^{\alpha}_k T_k}{T_k-t}
\,P_{tT_k}\,{\rm Cov}\left[\left. D_{T_k}, X^{\alpha}_{T_k}
\right|{\mathcal F}_t\right]\rd W^{\alpha k}_t \nonumber \\ && +
\sum_{k=1}^n D_{T_k}\rd {\mathds 1}_{\{t<T_k\}},
\end{eqnarray}
where $D_{T_k}=\Delta_{T_k}(X^{\alpha}_{T_1}, X^{\alpha}_{T_2},
\cdots, X^{\alpha}_{T_k})$ is the dividend at time $T_k$
$(k=1,2,\ldots,n)$. } \vspace{0.4cm}

\section{Time-dependent information flow}
\label{sec:8}

We consider now a generalisation of the foregoing material to the
situation in which the information-flow rate varies in time. The
time-dependent problem is of relevance to many circumstances. For
example, there will typically be more activity in a market during
the day than at night---such a consideration is important for
short-term investments. Alternatively, it may be that the annual
report of a firm is going to be published on a specified day---in
this case much more information concerning the future of the firm
may be made available on that day than normal.

We begin our analysis of the time-dependent case by considering
the situation where there is a single cash flow $D_T$ occurring at
$T$, and the associated market factor is the cash flow itself. In
this way we can focus our attention on mathematical issues arising
from the time dependence of the information flow rate. Once these
issues have been dealt with, we shall consider more complicated
cash-flow structures. For the market information process we
propose an expression of the form
\begin{eqnarray}
\xi_t = D_T \int_0^t \sigma_s \rd s +\beta_{tT}, \label{eq:2.2}
\end{eqnarray}
where the function $\{\sigma_s\}_{o\le s\le T}$ is taken to be
deterministic and nonnegative. We assume that $0<\int_0^T \sigma_s^2
ds<\infty$. The price process $\{S_t\}$ of the asset is then given
by
\begin{eqnarray}
S_t={\mathds 1}_{\{t<T\}} P_{tT}{\mathbb E}
\left[D_T\left|{\mathcal F}_t\right.\right]. \label{eq:2.1}
\end{eqnarray}
where the market filtration is, as in the previous sections,
assumed to be generated by the information process.

Our first task is to work out the conditional expectation in
(\ref{eq:2.1}). This can be achieved by use of a change-of-measure
technique, which will be outlined in Section~\ref{sec:9}. It will
be useful, however, to state the result first. We define the
conditional probability density process $\{\pi_t(x)\}$ by setting
\begin{eqnarray}
\pi_t(x) = \frac{\rd}{\rd x}\,{\mathbb Q}\left(\left. D_T\leq x
\right| {\mathcal F}_t \right).
\end{eqnarray}
The following result is obtained:

\vspace{0.4cm} \noindent {\bf Proposition~3}. {\it Let the
information process $\{\xi_t\}$ be given by {\rm (\ref{eq:2.2})}.
Then the conditional probability density process $\{\pi_t(x)\}$
for the random variable $D_T$ is given by
\begin{eqnarray}
\pi_t(x) = \frac{p(x)\,\re^{ x \left(\frac{1}{T-t}\,\xi_t \int_0^t
\sigma_s{\rm d}s+ \int_0^t \sigma_s{\rm d}\xi_s\right)
-\frac{1}{2} x^2 \left( \frac{1}{T-t}\left(\int_0^t \sigma_s{\rm
d} s \right)^2+ \int_0^t \sigma_s^2{\rm d}s\right)}}
{\int_{0}^\infty p(x)\,\re^{ x \left( \frac{1}{T-t}\,
\xi_t\int_0^t\sigma_s{\rm d}s+ \int_0^t \sigma_s{\rm d}\xi_s
\right) -\frac{1}{2} x^2 \left( \frac{1}{T-t}\left(\int_0^t
\sigma_s{\rm d} s\right)^2+ \int_0^t \sigma_s^2{\rm d}s\right)}\rd
x}. \label{eq:3.13}
\end{eqnarray}
} \vspace{0.4cm}

We deduce at once from Proposition 3 that the conditional
expectation of the random variable $D_T$ is
\begin{eqnarray}
D_{tT} = \frac{\int_{0}^\infty x p(x)\,\re^{ x \left(
\frac{1}{T-t}\,\xi_t\int_0^t\sigma_s{\rm d}s+ \int_0^t \sigma_s
{\rm d}\xi_s\right) -\frac{1}{2} x^2 \left(
\frac{1}{T-t}\left(\int_0^t \sigma_s{\rm d}s\right)^2+ \int_0^t
\sigma_s^2{\rm d}s\right)} {\rm d}x}{\int_{0}^\infty p(x)\,\re^{ x
\left( \frac{1}{T-t}\, \xi_t\int_0^t\sigma_s{\rm d}s+ \int_0^t
\sigma_s {\rm d}\xi_s\right) -\frac{1}{2} x^2 \left(
\frac{1}{T-t}\left(\int_0^t \sigma_s{\rm d}s\right)^2+ \int_0^t
\sigma_s^2{\rm d}s\right)}\rd x}. \label{eq:3.14}
\end{eqnarray}
The associated price process $\{S_t\}$ is then given by $S_t=
{\mathds 1}_{\{t<T\}}P_{tT}D_{tT}$.

\section{Changes of measure for Brownian bridges}
\label{sec:9}

Since the information process is a Brownian bridge with a random
drift, we shall require formulae relating a Brownian bridge with
drift in one measure to a standard Brownian bridge in another
measure to establish Proposition~3 . We proceed as follows. First
we recall a well-known integral representation for the Brownian
bridge. Let the probability space $(\Omega,{\mathcal F},{\mathbb
Q})$ be given, with a filtration $\{{\mathcal G}_t\}_{0\le
t<\infty}$, and let $\{B_t\}$ be a standard $\{{\mathcal
G}_t\}$-Brownian motion. Then the process $\{\beta_{tT}\}$,
defined by
\begin{eqnarray}
\beta_{tT} = (T-t)\int_0^t \frac{1}{T-s}\, \rd B_s,
\label{eq:3.1}
\end{eqnarray}
for $0\leq t<T$, and by $\beta_{tT}=0$ for $t=T$, is a standard
Brownian bridge over the interval $[0,T]$. Expression (\ref{eq:3.1})
converges to zero as $t\to T$; see, e.g., Karatzas and Shreve
\cite{karatzas}, Protter \cite{protter}). The filtration
$\{{\mathcal G}_t\}$ is larger than the market filtration
$\{{\mathcal F}_t\}$. In particular, since $\{\beta_{tT}\}$ is
adapted to $\{{\mathcal G}_t\}$ we can think of $\{{\mathcal G}_t\}$
as the filtration describing the information available to an
``insider'' who can distinguish between what is noise and what is
not.

Let $D_T$ be a random variable on $(\Omega,{\mathcal F},{\mathbb
Q})$. We assume that $D_T$ is ${\mathcal G}_0$-measurable and that
$D_T$ is independent of $\{\beta_{tT}\}$. Thus the value of $D_T$ is
known ``all along'' to the insider, but not to the typical market
participant. For simplicity in what follows we assume that $D_T$ is
bounded; this condition can be relaxed with the introduction of an
appropriate Novikov-type condition; but we will not pursue the more
general situation here. Define the deterministic nonnegative process
$\{\nu_t\}_{0\leq t\leq T}$ by
\begin{eqnarray}
\nu_t = \sigma_t + \frac{1}{T-t} \int_0^t \sigma_s \rd s,
\label{eq:3.4}
\end{eqnarray}
and let $\{\xi_t\}$ be defined as in (\ref{eq:2.2}). We define the
process $\{\Lambda_t\}_{0\leq t<T}$ by the relation
\begin{eqnarray}
\frac{1}{\Lambda_t} = \exp\left( -D_T\int_0^t\nu_s\rd B_{s}-\half
D_T^2 \int_0^t \nu_s^2\rd s\right). \label{eq:3.3}
\end{eqnarray}
With these elements in hand, we fix a time horizon $U\in(0,T)$ and
introduce a probability measure ${\mathbb B}$ on ${\mathcal
G}_{U}$ by the relation
\begin{eqnarray}
\rd{\mathbb B}=\Lambda_{U}^{-1}\rd{\mathbb Q}.
\end{eqnarray}
Then we have the following facts: (i) The process
$\{W_t^*\}_{0\leq t<U}$ defined by
\begin{eqnarray}
W_t^* = D_T\int_0^t \nu_s\rd s + B_t \label{eq:3.5}
\end{eqnarray}
is a ${\mathbb B}$-Brownian motion. (ii) The process $\{\xi_t\}$
defined by (\ref{eq:2.2}) is a ${\mathbb B}$-Brownian bridge and
is independent of $D_T$. (iii) The random variable $D_T$ has the
same probability law with respect to ${\mathbb B}$ and ${\mathbb
Q}$. (iv) The conditional expectation for any integrable function
$f(D_T)$ of the random variable $D_T$ can be expressed in the form
\begin{eqnarray}
{\mathbb E}^{\mathbb Q} [f(D_T)|{\mathcal F}_t^\xi] =
\frac{{\mathbb E}^{{\mathbb B}}\left[f(D_T)\Lambda_t\big|
{\mathcal F}_t^\xi\right]}{{\mathbb E}^{{\mathbb B}}
\left[\Lambda_t \big| {\mathcal F}_t^\xi\right]}. \label{eq:3.6}
\end{eqnarray}
We note that the measure ${\mathbb B}$ is independent of the
specific choice of the time horizon $U$ in the sense that if
${\mathbb B}$ is defined on ${\mathcal G}_{U'}$ for some $U'>U$,
then the restriction of that measure to ${\mathcal G}_{U}$ agrees
with the measure ${\mathbb B}$ as already defined.

When we say that $\{\xi_t\}$ is a ${\mathbb B}$-Brownian bridge what
we mean, more precisely, is that $\xi_0=0$, that $\{\xi_t\}$ is
${\mathbb B}$-Gaussian, that ${\mathbb E}^{{\mathbb B}}[\xi_t]=0$,
and that
\begin{eqnarray}
{\mathbb E}^{{\mathbb B}}[\xi_s\xi_t]=\frac{s(T-t)}{T}
\end{eqnarray}
for $0\le s\le t\le U$. Thus with respect to the measure ${\mathbb
B}$ the process $\{\xi_t\}_{0\le t\le U}$ has the properties of a
standard $[0,T]$-Brownian bridge that has been truncated at time
$U$.  The fact that $\{\xi_t\}$ is a ${\mathbb B}$-Brownian bridge
can be verified as follows. By (\ref{eq:2.2}), (\ref{eq:3.1}), and
(\ref{eq:3.5}) we have
\begin{eqnarray}
\xi_t &=& D_T \int_0^t \sigma_s \rd s + (T-t) \int_0^t
\frac{1}{T-s}\, \rd B_s \nonumber \\ &=& D_T \int_0^t \sigma_s \rd
s + (T-t) \int_0^t \frac{1}{T-s}\,(\rd W_s^*-D_T\nu_s\rd s)
\nonumber \\ &=& D_T\left(\int_0^t \sigma_s \rd s-(T-t)\int_0^t
\frac{1}{T-s}\,\nu_s\rd s\right) + (T-t)\int_0^t \frac{1}{T-s}\,
\rd W_s^* \nonumber \\ &=& (T-t)\int_0^t \frac{1}{T-s}\, \rd
W_s^*, \label{eq:3.7}
\end{eqnarray}
where in the final step we used the relation
\begin{eqnarray}
\int_0^t \frac{1}{T-s}\,\nu_s\rd s = \frac{1}{T-t}\int_0^t
\sigma_s \rd s. \label{eq:3.8}
\end{eqnarray}
This formula can be verified explicitly by differentiation, which
then gives us (\ref{eq:3.4}). In (\ref{eq:3.7}) we see that
$\{\xi_t\}$ has been given the standard integral representation of a
Brownian bridge.

We remark, incidentally, that (\ref{eq:3.6}) can be thought of a
variation of the Kallianpur-Striebel formula appearing in the
literature of nonlinear filtering (see, for example, Bucy and Joseph
\cite{bj}, Davis and Marcus \cite{davis}, Fujisaki \textit{et al}.
\cite{fujisaki}, Kallianpur and Striebel \cite{ks}, Krishnan
\cite{krishnan}, and Liptser and Shiryaev \cite{ls}).

\section{Derivation of the conditional density}
\label{sec:10}

We have introduced the idea of measure changes associated with
Brownian bridges in order to introduce formula (\ref{eq:3.6}), which
involves the density process $\{\Lambda_t\}$, which in
(\ref{eq:3.3}) is defined in terms of the ${\mathbb Q}$-Brownian
motion $\{B_t\}$. On the other hand, the expectations in
(\ref{eq:3.6}) are conditional with respect to the information
generated by $\{\xi_t\}$. Therefore, it will be convenient to
express $\{\Lambda_t\}$ in terms of $\{\xi_t\}$. To do this we
substitute (\ref{eq:3.5}) in (\ref{eq:3.3}) to obtain
\begin{eqnarray}
\Lambda_t = \exp\left( D_T\int_0^t\nu_s\rd W_s^*-\half D_T^2
\int_0^t \nu_s^2\rd s\right). \label{eq:3.90}
\end{eqnarray}
We then observe, by differentiating (\ref{eq:3.7}), that
\begin{eqnarray}
\rd \xi_t = -\frac{\xi_t}{T-t}\,\rd t + \rd W_t^*. \label{eq:xi.1}
\end{eqnarray}
Substituting this relation in (\ref{eq:3.90}) we obtain
\begin{eqnarray}
\Lambda_t = \exp\left[ D_T\left( \int_0^t\nu_s\rd\xi_s+\int_0^t
\frac{1}{T-s}\,\nu_s\xi_s\rd s\right) - \half D_T^2 \int_0^t
\nu_s^2 \rd s\right] . \label{eq:3.9}
\end{eqnarray}

In principle at this point all we need to do is to substitute
(\ref{eq:3.4}) into (\ref{eq:3.9}) to obtain the result for
$\{\Lambda_t\}$. In practice, further simplification can be
achieved. To this end, we note that by taking the differential of
the coefficient of $D_T$ in the exponent of (\ref{eq:3.9}) we get
\begin{eqnarray}
\rd\left( \int_0^t\nu_s\rd\xi_s+\int_0^t \frac{1}{T-s}\,\nu_s\xi_s
\rd s\right)&=&\nu_t\left(\rd\xi_t+\frac{1}{T-t}\,\xi_t\rd t
\right) \nonumber \\ &=& \left(\sigma_t + \frac{1}{T-t} \int_0^t
\sigma_s \rd s \right)\left(\rd\xi_t+ \frac{1}{T-t}\,\xi_t\rd
t\right) \nonumber \\ &=& \rd \left(
\frac{1}{T-t}\,\xi_t\int_0^t\sigma_s \rd s +
\int_0^t\sigma_s\rd\xi_s\right) . \label{eq:3.10}
\end{eqnarray}
Then integrating both sides of (\ref{eq:3.10}) we obtain:
\begin{eqnarray}
\int_0^t\nu_s\rd\xi_s+\int_0^t \frac{1}{T-s}\,\nu_s\xi_s \rd s =
\frac{1}{T-t}\,\xi_t\int_0^t\sigma_s \rd s +
\int_0^t\sigma_s\rd\xi_s . \label{eq:3.105}
\end{eqnarray}
Similarly, by taking the differential of the coefficient of
$-\frac{1}{2}D_T^2$ in the exponent of (\ref{eq:3.9}) and making use
of (\ref{eq:3.4}), we find
\begin{eqnarray}
\nu_t^2 \rd t &=& \left[\sigma_t^2+ 2\frac{1}{T-t}\,\sigma_t
\int_0^t \sigma_s \rd s + \frac{1}{(T-t)^2} \left( \int_0^t
\sigma_s\rd s\right)^2 \right]\rd t \nonumber \\ &=& \rd \left[
\frac{1}{T-t} \left( \int_0^t \sigma_s\rd s\right)^2+\int_0^t
\sigma_s^2\rd s\right]. \label{eq:3.11}
\end{eqnarray}
Therefore, by integrating both sides of (\ref{eq:3.11}) we obtain
an identity for the coefficient of $-\frac{1}{2}D_T^2$.

It follows by virtue of the two identities just obtained that
$\{\Lambda_t\}$ can be expressed in terms of $\{\xi_t\}$. More
explicitly, we have
\begin{eqnarray}
\Lambda_t={\textstyle\exp\left[D_T\left(\frac{1}{T-t}\, \xi_t
\int_0^t \sigma_s\rd s+ \int_0^t \sigma_s \rd \xi_s\right) -\half
D_T^2 \left( \frac{1}{T-t}\left(\int_0^t \sigma_s\rd s\right)^2+
\int_{0}^{t} \sigma_{s}^2\rd s\right) \right]}. \label{eq:3.12}
\end{eqnarray}
Note that by transforming (\ref{eq:3.9}) into (\ref{eq:3.12}) we
have eliminated a term having $\{\xi_t\}$ in the integrand, thus
achieving a considerable simplification. Proposition~3 can then be
deduced if we use equation (\ref{eq:3.8}) and the basic relation
\begin{eqnarray}
{\mathbb Q}\left( \left. D_T\leq x\right|{\mathcal F}_t\right) =
{\mathbb E}^{\mathbb Q}\left[ \left. {\mathds 1}_{\{D_T\leq
x\}}\right|{\mathcal F}_t \right]. \label{eq:QQ}
\end{eqnarray}
In particular, since $D_T$ and $\{\xi_t\}$ are independent under
the bridge measure, by virtue of (\ref{eq:3.6}), (\ref{eq:3.12}),
and (\ref{eq:QQ}) we obtain
\begin{eqnarray}
{\mathbb Q}\left( \left. D_T\leq x\right|{\mathcal F}_t \right) =
\frac{\int_0^x p(y)\,\re^{ y\left(\frac{1}{T-t}\,\xi_t \int_0^t
\sigma_s{\rm d}s+ \int_0^t \sigma_s{\rm d}\xi_s\right)
-\frac{1}{2} y^2 \left( \frac{1}{T-t}\left(\int_0^t \sigma_s{\rm
d} s \right)^2+ \int_0^t \sigma_s^2{\rm d}s\right)}{\rm d}y}
{\int_{0}^\infty p(y)\,\re^{ y\left( \frac{1}{T-t}\,
\xi_t\int_0^t\sigma_s{\rm d}s+ \int_0^t \sigma_s{\rm d}\xi_s
\right) -\frac{1}{2} y^2 \left( \frac{1}{T-t}\left(\int_0^t
\sigma_s{\rm d} s\right)^2+ \int_0^t \sigma_s^2{\rm d}s\right)}\rd
y}, \label{eq:3.135}
\end{eqnarray}
from which we immediately infer Proposition~3 by differentiation
with respect to $x$.

We conclude this section by noting that an alternative expression
for $\{\pi_t(x)\}$, written in terms of $\{W_t^*\}$, is given by
\begin{eqnarray}
\pi_t(x) = \frac{p(x)\,\exp\left( x \int_0^t \nu_u{\rm d}W_u^*-
\half x^2 \int_0^t \nu_u^2{\rm d}u\right)} {\int_{0}^\infty
p(x)\,\exp\left( x \int_0^t \nu_u{\rm d}W_u^*- \half x^2 \int_0^t
\nu_u^2{\rm d}u\right)\rd x} . \label{eq:3.150}
\end{eqnarray}

\section{Consistency relations}
\label{sec:11}

Before we proceed to analyse in detail the dynamics of the price
process $\{S_t\}$, first we shall establish a useful
\textit{dynamical consistency} relation satisfied by prices obtained
in the information-based framework. By ``consistency'' we have in
mind the following. Suppose that we re-initialise the information
process at an intermediate time $s\in(0,T)$ by specifying the value
$\xi_s$ of the information at that time. For the framework to be
dynamically consistent, we require that the remainder of the period
$[s,T]$ admits a representation in terms of a suitably
``renormalised'' information process. Specifically, we have:

\vspace{0.4cm} \noindent {\bf Proposition~4}. {\it Let $0\leq
s\leq t \leq T$. The conditional probability $\pi_t(x)$ can be
written in terms of the intermediate conditional probability
$\pi_s(x)$ in the form
\begin{eqnarray}
\pi_t(x) = \frac{\pi_s(x)\,\re^{ x \left(\frac{1}{T-t}\,\eta_t
\int_s^t {\tilde\sigma}_u{\rm d}u+\int_s^t{\tilde\sigma}_u{\rm d}
\eta_u\right) -\frac{1}{2} x^2 \left( \frac{1}{T-t}\left(\int_s^t
{\tilde\sigma}_u{\rm d}u \right)^2+ \int_s^t {\tilde\sigma}_u^2
{\rm d}u\right)}} {\int_{0}^\infty \pi_s(x)\,\re^{ x
\left(\frac{1}{T-t}\,\eta_t \int_s^t {\tilde\sigma}_u{\rm
d}u+\int_s^t{\tilde\sigma}_u{\rm d} \eta_u\right) -\frac{1}{2} x^2
\left( \frac{1}{T-t}\left(\int_s^t {\tilde\sigma}_u{\rm d}u
\right)^2+ \int_s^t {\tilde\sigma}_u^2 {\rm d}u\right)}\rd x},
\label{eq:x.1}
\end{eqnarray}
where
\begin{eqnarray}
{\tilde\sigma}_u = \sigma_u + \frac{1}{T-s}\int_0^s \sigma_v {\rm
d}v \label{eq:x.2}
\end{eqnarray}
is the re-initialised market information-flow rate, and
\begin{eqnarray}
\eta_t = \xi_t - \frac{T-t}{T-s}\,\xi_s \label{eq:x.3}
\end{eqnarray}
is the re-initialised information process. } \vspace{0.4cm}

The fact that $\{\eta_t\}_{s\leq t\leq T}$ represents the updated
information process bridging the interval $[s,T]$ can be seen as
follows. First we note that $\eta_s=0$ and that $\eta_T=\xi_T$.
Substituting (\ref{eq:2.2}) in (\ref{eq:x.3}) we find that
\begin{eqnarray}
\eta_t = D_T \int_s^t {\tilde\sigma}_u{\rm d}u + \gamma_{tT},
\label{eq:x.4}
\end{eqnarray}
where ${\tilde\sigma}_u$ is as defined in (\ref{eq:x.2}), and
\begin{eqnarray}
\gamma_{tT}= \beta_{tT} - \frac{T-t}{T-s}\,\beta_{sT} .
\label{eq:x.5}
\end{eqnarray}
A calculation making use of the covariance of the Brownian bridge
$\{\beta_{tT}\}$ shows that the Gaussian process
$\{\gamma_{tT}\}_{s\leq t\leq T}$ is a standard Brownian bridge
over the interval $[s,T]$. It thus follows that $\{\eta_t\}$ is
the information bridge interpolating the interval $[s,T]$.

To verify (\ref{eq:x.1}) we note that (\ref{eq:3.150}) can be
written in the form
\begin{eqnarray}
\pi_t(x) = \frac{\pi_s(x)\,\exp\left( x \int_s^t \nu_u{\rm d}
W_u^*- \half x^2 \int_s^t \nu_u^2{\rm d}u\right)}{\int_{0}^\infty
\pi_s(x)\,\exp\left( x \int_s^t \nu_u{\rm d}W_u^*- \half x^2
\int_s^t \nu_u^2{\rm d}u\right)\rd x} . \label{eq:x.6}
\end{eqnarray}
The identity (\ref{eq:3.10}) then implies that
\begin{eqnarray}
\int_s^t \nu_u{\rm d} W_u^* &=& \frac{1}{T-t}\,\xi_t \int_s^t
\sigma_u {\rm d}u + \int_s^t \sigma_u {\rm d}\xi_u + \Big( \frac{
\xi_t}{T-t}-\frac{\xi_s}{T-s}\Big)\int_0^s\sigma_u{\rm d}u
\nonumber \\ &=& \frac{1}{T-t}\,\eta_t \int_s^t {\tilde\sigma}_u
{\rm d}u+\int_s^t{\tilde\sigma}_u{\rm d} \eta_u, \label{eq:x.7}
\end{eqnarray}
where we have made use of (\ref{eq:x.2}) and (\ref{eq:x.3}).
Similarly, (\ref{eq:3.105}) implies that
\begin{eqnarray}
\int_s^t \nu_u^2{\rm d}u &=& \frac{1}{T-t} \left( \int_0^t
\sigma_u{\rm d}u\right)^2 - \frac{1}{T-s} \left( \int_0^s \sigma_u
{\rm d}u\right)^2 + \int_s^t \sigma_u^2 {\rm d}u \nonumber \\ &=&
\frac{1}{T-t}\left(\int_s^t {\tilde\sigma}_u{\rm d}u \right)^2+
\int_s^t {\tilde\sigma}_u^2 {\rm d}u . \label{eq:x.8}
\end{eqnarray}
Substitution of (\ref{eq:x.7}) and (\ref{eq:x.8}) into
(\ref{eq:x.6}) establishes (\ref{eq:x.1}). In particular, the form
of (\ref{eq:x.1}) is identical to the original formula
(\ref{eq:3.13}), modulo the indicated renormalisation of the
information process and the associated information flow rate.

\section{Expected dividend}

\label{sec:12} The goal of sections \ref{sec:8}, \ref{sec:9}, and
\ref{sec:10} was to obtain an expression for the conditional
expectation (\ref{eq:3.13a}) in the case of a single-dividend asset
in the case of a time-dependent information-flow rate. In the
analysis of the associated price process it will therefore be useful
to work out the dynamics of the conditional expectation of the
dividend. In particular, an application of Ito's rule to
(\ref{eq:3.14}), after some rearrangement, shows that
\begin{eqnarray}
\rd D_{tT} = \nu_t V_t \left( \frac{1}{T-t}\,\xi_t-\nu_t
D_{tT}\right)\rd t + \nu_t V_t \rd \xi_t, \label{eq:4.1}
\end{eqnarray}
where $\{V_t\}$ is the conditional variance of the random variable
$D_T$, given by (\ref{eq:2.16}). Let us define a new process
$\{W_t\}$ by
\begin{eqnarray}
W_t = \xi_t + \int_0^t \frac{1}{T-s}\,\xi_s\rd s - \int_0^t \nu_s
D_{sT} \rd s. \label{eq:4.3}
\end{eqnarray}
We refer to $\{W_t\}$ as the ``innovation process''. It follows from
the definition of $\{W_t\}$ that
\begin{eqnarray}
\rd D_{tT}= \nu_t V_t\, \rd W_t. \label{eq:4.4}
\end{eqnarray}
Since $\{D_{tT}\}$ is an $\{{\mathcal F}_t\}$-martingale we are thus
led to conjecture that $\{W_t\}$ must also be an $\{{\mathcal
F}_t\}$-martingale. In fact, we have the following result:

\vspace{0.4cm} \noindent {\bf Proposition~5}. {\it The process
$\{W_t\}$ defined by {\rm (\ref{eq:4.3})} is an $\{{\mathcal
F}_t\}$-Brownian motion under ${\mathbb Q}$}. \vspace{0.4cm}

Proof. We need to establish that (i) $\{W_t\}$ is an $\{{\mathcal
F}_t\}$-martingale, and that (ii) $(\rd W_t)^2=\rd t$. Writing
${\mathbb E}_t[-]={\mathbb E}^{\mathbb Q}[-|{\mathcal F}_t]$ and
letting $t\leq u$ we have
\begin{eqnarray}
{\mathbb E}_t\left[W_u\right]= {\mathbb E}_t\left[\xi_u\right] +
{\mathbb E}_t\left[\int_0^u\frac{1}{T-s}\,\xi_s\rd s\right] -
{\mathbb E}_t\left[\int_0^u\!\nu_s D_{sT} \rd s\right] .
\label{eq:4.5}
\end{eqnarray}
Splitting the second two terms on the right into integrals between
$0$ and $t$, and between $t$ and $u$, we obtain
\begin{eqnarray}
{\mathbb E}_t\left[W_u\right]&=&{\mathbb E}_t[\xi_u] +
\int_0^t\frac{1}{T-s}\,\xi_s\rd s - \int_0^t\!\nu_s D_{sT} \rd s
\nonumber \\ && + \int_t^u\frac{1}{T-s} \, {\mathbb E}_t[\xi_s]\rd
s - \int_t^u\!\nu_s {\mathbb E}_t[D_{sT}] \rd s. \label{eq:4.55}
\end{eqnarray}
The martingale property of the conditional expectation implies
that ${\mathbb E}_t[D_{sT}]=D_{tT}$ for $t\leq s$, which allows us
to simplify the last term. To simplify the expression for the
expectation ${\mathbb E}_t[\xi_s]$ for $t\leq s$ we use the tower
property:
\begin{eqnarray}
{\mathbb E}_t[\beta_{sT}]={\mathbb E}_t [{\mathbb E}
[\beta_{sT}\vert H_T,\beta_{tT}]]={\mathbb E}_t [{\mathbb E}
[\beta_{sT}\vert\beta_{tT}]]. \label{eq:24e}
\end{eqnarray}
To calculate the inner expectation ${\mathbb E}
[\beta_{sT}\vert\beta_{tT}]$ here we use the fact that the random
variable $\beta_{sT}/(T-s)-\beta_{tT}/(T-t)$ is independent of
$\beta_{tT}$ and deduce that
\begin{eqnarray}
{\mathbb E}[\beta_{sT}|\beta_{tT}]=\frac{T-s}{T-t}\,\beta_{tT},
\label{eq:24g}
\end{eqnarray}
from which it follows that
\begin{eqnarray}
{\mathbb E}_t\left[\beta_{sT}\right] = \frac{T-s}{T-t}\, {\mathbb
E}_t[\beta_{tT}].
\end{eqnarray}
As a result we obtain
\begin{eqnarray}
{\mathbb E}_t[\xi_s]=D_{tT}\int_0^s \sigma_v \rd v +
\frac{T-s}{T-t}\, {\mathbb E}_t[\beta_{tT}]. \label{eq:aa}
\end{eqnarray}
We also recall the definition of $\{W_t\}$ given by
(\ref{eq:4.3}), which implies that
\begin{eqnarray}
\int_0^t\frac{1}{T-s}\,\xi_s\rd s - \int_0^t\!\nu_s D_{sT} \rd s =
W_t-\xi_t. \label{eq:bb}
\end{eqnarray}
Therefore, substituting (\ref{eq:aa}) and (\ref{eq:bb}) into
(\ref{eq:4.55}) we obtain
\begin{eqnarray}
{\mathbb E}_t\left[W_u\right]&=& D_{tT} \int_0^u \sigma_s\rd s +
W_t - \xi_t + D_{tT} \int_t^u\frac{1}{T-s}\Big(\int_0^s\sigma_v\rd
v\Big) \rd s - D_{tT} \int_t^u\nu_s\rd s \nonumber \\ && +
{\mathbb E}_t[\beta_{tT}]. \label{eq:4.8}
\end{eqnarray}
Next we split the first term into an integral from $0$ to $t$ and
an integral from $t$ to $u$, and we insert the definition
(\ref{eq:3.4}) of $\{\nu_t\}$ into the fifth term. The result is:
\begin{eqnarray}
{\mathbb E}_t\left[W_u\right]&=&W_t+ D_{tT}\int_0^t\sigma_s \rd s
+ {\mathbb E}_t [\beta_{tT}]-\xi_t. \label{eq:4.85}
\end{eqnarray}
Finally, if we make use of the fact that $\xi_t={\mathbb E}_t
[\xi_t]$, and hence that
\begin{eqnarray}
\xi_t = D_{tT} \int_0^t\sigma_s \rd s + {\mathbb E}_t[\beta_{tT}],
\label{eq:4.9}
\end{eqnarray}
we see that $\{W_t\}$ satisfies the martingale condition. On the
other hand, by virtue of (\ref{eq:4.3}) we have $(\rd W_t)^2=\rd
t$. We conclude that $\{W_t\}$ is an $\{{\mathcal F}_t\}$-Brownian
motion under ${\mathbb Q}$. \hspace*{\fill} $\square$

\section{Asset prices and derivative prices}
\label{sec:14}

We are now in a position to consider in more detail the dynamics
of the price process of an asset paying a single dividend $D_T$ in
the case of a time-dependent information flow. For $\{S_t\}$ we
have $S_t= {\mathds 1}_{\{t<T\}}P_{tT}D_{tT}$, or equivalently
\begin{eqnarray}
S_{t} = {\mathds 1}_{\{t<T\}}P_{tT} \frac{\int_{0}^\infty x
p(x)\,\re^{ x \left(\frac{1}{T-t}\xi_t\int_0^t\sigma_s{\rm d} s+
\int_0^t \sigma_s {\rm d}\xi_s\right) -\frac{1}{2} x^2 \left(
\frac{1}{T-t} \left( \int_0^t \sigma_s{\rm d}s\right)^2+ \int_0^t
\sigma_s^2{\rm d} s\right)}{\rm d}x}{\int_{0}^\infty p(x)\,\re^{x
\left(\frac{1}{T-t}\xi_t\int_0^t\sigma_s{\rm d}s+ \int_0^t
\sigma_s {\rm d}\xi_s\right) -\frac{1}{2} x^2 \left(
\frac{1}{T-t}\left( \int_0^t \sigma_s{\rm d}s\right)^2+ \int_0^t
\sigma_s^2\rd s \right)}\rd x}. \label{eq:7.1}
\end{eqnarray}
A calculation making use of (\ref{eq:4.4}) shows that for the
dynamics of the price we have
\begin{eqnarray}
\rd S_{t} = r_t S_{t}\rd t + \Gamma_{tT} \rd W_t, \label{eq:7.2}
\end{eqnarray}
where the asset price volatility is given by $\Gamma_{tT} = \nu_t
P_{tT} V_t$, where $V_t$ is the conditional variance of the
dividend, given by (\ref{eq:2.16}). It should be evident by virtue
of its definition that $\{V_t\}$ is a supermartingale. More
specifically, for the dynamics of $\{V_t\}$ we obtain
\begin{eqnarray}
\rd V_t = -\nu_t^2 V_t^2\rd t + \nu_t\kappa_t\rd W_t,
\label{eq:5.2}
\end{eqnarray}
where $\kappa_t$ denotes the third conditional moment of $D_T$,
given by $\kappa_t = {\mathbb E}_t\left[ (D_T-D_{tT})^3 \right]$.

Although we have derived (\ref{eq:7.1}) by assuming that the price
process is induced by the market information $\{\xi_t\}$, the
result to be shown in Section~\ref{sec:15} demonstrates that we
can regard the dynamical equation (\ref{eq:7.2}) for the price
process as given, and then deduce the structure of the underlying
information. The information-based interpretation of the modelling
framework, however, is more appealing. According to this
interpretation there is a flow of market information, which is
available to all market participants and is represented by the
filtration generated by $\{\xi_t\}$. Given this information, each
participant will ``act'', in our interpretation, so as to minimise
the risk adjusted future P\&L variance associated with the cash
flow under consideration. The future P\&L is determined by the
value of $D_T$, and the estimate of $D_T$ that minimises its
variance is indeed given by the conditional expectation
(\ref{eq:3.13a}). By discounting this expectation with $P_{tT}$ we
recover the price process $\{S_{t}\}$.

We note that $\{\Gamma_{tT}\}$ is  ``infinitely stochastic'' in
the sense that all of the higher-order volatilities (the
volatility of the volatility, and so on) are stochastic. These
higher-order volatilities have a natural interpretation: the
volatility of the asset price is determined by the variance of the
random cash flow; the volatility of the volatility is determined
by the skewness of $D_T$; its volatility is determined by the
kurtosis of $D_T$; and so on.

The fact that the asset price in the bridge measure is given by a
function of a Gaussian random variable means that the pricing of
derivatives is numerically straightforward. We have seen this in the
case of a constant information-flow rate, but the result holds in
the time-dependent case as well. For example, consider a
European-style call option with strike $K$ and maturity $t$, where
$t\leq T$, for which the value is (\ref{Eopt}). If we express the
asset price $S_t$ on the option maturity date in terms of the
${\mathbb B}$-Brownian motion $\{W_t^*\}$ we find
\begin{eqnarray}
C_0 = P_{0t} {\mathbb E}^{{\mathbb Q}}\left[ \frac{1}{\Phi_t}
\left\{\int_{0}^\infty ( P_{tT}x-K) p(x)\, \exp
\left(x\int_0^t\nu_s{\rm d}W_s^*-\half x^2\int_0^t\nu_s^2{\rm
d}s\right)  \rd x\right\}^+ \right], \label{eq:7.4.0}
\end{eqnarray}
where
\begin{eqnarray}
\Phi_t = \int_{0}^\infty p(x)\, \exp \left(x\int_0^t\nu_s{\rm d}
W_s^*-\half x^2\int_0^t\nu_s^2{\rm d}s\right)  \rd x.
\label{eq:7.4.1}
\end{eqnarray}

To proceed we shall use the factor $1/\Phi_t$ in (\ref{eq:7.4.0}) to
make a change of measure on $(\Omega, {\mathcal F}_t)$. The idea is
as follows. We fix a time horizon $u$ at or beyond the option
expiration but before the bond maturity, so $t\leq u<T$, and define
a process $\{\Phi_t\}_{0\leq t\leq u}$ by use of the expression
(\ref{eq:7.4.1}), where now we let $t$ vary in the range $[0,u]$. By
an application of Ito calculus on (\ref{eq:7.4.1}) we see that $\rd
\Phi_t = \nu_t D_{tT}\Phi_t \,\rd W^*_t$. On the other hand, it
follows from (\ref{eq:xi.1}) and (\ref{eq:4.3}) that the ${\mathbb
B}$-Brownian motion $\{W_t^*\}$ and the ${\mathbb Q}$-Brownian
motion $\{W_t\}$ are related by
\begin{eqnarray}
\rd W_t^* = \rd W_t + \nu_t D_{tT} \rd t.
\end{eqnarray}
Therefore, in terms of $\{W_t\}$ we have
\begin{eqnarray}
\rd \Phi_t = \nu_t^2 D_{tT}^2 \Phi_t \rd t + \nu_t D_{tT}\Phi_t \,
\rd W_t, \label{eq:7.4.2}
\end{eqnarray}
from which it follows that
\begin{eqnarray}
\rd \Phi_t^{-1} = - \nu_t D_{tT}\Phi_t^{-1}\rd W_t .
\label{eq:8.105}
\end{eqnarray}
Upon integration we deduce
\begin{eqnarray}
\Phi_t^{-1} = \exp\left( -\int_0^t \nu_s D_{sT}\,\rd W_s -\half
\int_0^t\nu_s^2 D_{sT}^2\,\rd s \right). \label{eq:8.106}
\end{eqnarray}
Since $\{\nu_s D_{sT}\}$ is bounded, and $s\leq u<T$, we see that
$\{\Phi_s^{-1}\}_{0\leq s\leq u}$ is a ${\mathbb Q}$-martingale with
${\mathbb E}^{\mathbb Q}[\Phi_t^{-1}]=1$, where $t$ is the option
maturity date. Therefore, the factor $1/\Phi_t$ in (\ref{eq:7.4.0})
can be used to effect a change of measure ${\mathbb Q}\to{\mathbb
B}$ on $(\Omega,{\mathcal F}_t)$. We note that while on the space
$(\Omega,{\mathcal G}_t)$ it is the process $\{\Lambda_t\}$
introduced in (\ref{eq:3.3}) that defines the measure change from
${\mathbb B}$ and ${\mathbb Q}$, on $(\Omega,{\mathcal F}_t)$ it is
$\Phi_t={\mathbb E}^{\mathbb Q} [\Lambda_t|{\mathcal F}_t]$ that
defines the relevant measure change. As a consequence, by changing
the measure in (\ref{eq:7.4.0}) we obtain
\begin{eqnarray}
C_0 = P_{0t} {\mathbb E}^{{\mathbb B}}\left[
\left\{\int_{0}^\infty ( P_{tT}x-K) p(x)\,
\exp\left(x\int_0^t\nu_s{\rm d}W_s^*-\half x^2\int_0^t\nu_s^2{\rm
d}s\right)  \rd x\right\}^+ \right]. \label{eq:7.4}
\end{eqnarray}
This result should be compared with (\ref{eq:3-3a}). We note that in
the bridge measure the expression $\int_0^t\nu_s{\rm d}W_s^*$ is a
Gaussian random variable with mean zero and variance
\begin{eqnarray}
\omega_t^2=\int_0^t\nu_s^2 \rd s = \frac{1}{T-t} \left( \int_0^t
\sigma_s\rd s\right)^2+\int_0^t \sigma_s^2\rd s .
\end{eqnarray}
Here we have used the relation (\ref{eq:3.11}). Therefore, if we set
\begin{eqnarray}
Y=\omega_t^{-1}\int_0^t\nu_s{\rm d}W_s^*,
\end{eqnarray}
it follows that $Y$ is ${\mathbb B}$-Gaussian. For the call price we
thus have
\begin{eqnarray}
C_0 = P_{0t} {\mathbb E}^{{\mathbb B}}\left[ \left\{
\int_{0}^\infty ( P_{tT}x-K) p(x)\re^{\omega_t xY-\frac{1}{2}
\omega_t^2 x^2} \rd x\right\}^+ \right],
\end{eqnarray}
and hence
\begin{eqnarray}
C_0= P_{0t}
\frac{1}{\sqrt{2\pi}} \int_{y=-\infty}^\infty\!\re^{-\frac{1}{2}
y^2} \left( \int_{x=0}^\infty\!(P_{tT}x-K) p(x)\re^{\omega_t
xy-\frac{1}{2} \omega_t^2 x^2} \rd x\right)^+ \rd y.
\label{eq:7.45}
\end{eqnarray}
We observe that there exists a critical value $y=y^*$ such that
the argument of the ``plus'' function vanishes in the expression
above. Thus $y^*$ is given by
\begin{eqnarray}
\int_{0}^\infty (P_{tT}x-K) p(x)\,\re^{\omega_t x y^* -\frac{1}{2}
\omega_t^2 x^2} \rd x=0. \label{eq:7.46}
\end{eqnarray}
As a consequence the call price can be written
\begin{eqnarray}
C_0 = P_{0t} \frac{1}{\sqrt{2\pi}} \int_{y=y^* }^\infty\!
\re^{-\frac{1}{2} y^2} \left( \int_{x=0}^\infty\!(P_{tT}x-K) p(x)
\re^{\omega_t xy-\frac{1}{2} \omega_t^2 x^2} \rd x\right) \rd y.
\label{eq:7.47}
\end{eqnarray}
The integration in the $y$ variable can be performed, and we
deduce the following representation for the call price:
\begin{eqnarray}
C_0 = P_{0t} \int_{0}^\infty (P_{tT}x-K) p(x) N(\omega_t x - y^* )
\rd x. \label{eq:7.48}
\end{eqnarray}
When the cash flow is represented by a discrete random variable
and the information-flow rate is constant, this result reduces to
an expression equivalent to the option pricing formula derived in
Brody \textit{et al}. \cite{bh1}. If the cash flow is a continuous
random variable and the information flow rate is constant then we
recover the expression (\ref{eq:ec1}) given in section IV (see
also Rutkowski and Yu \cite{rutkowski}).

We conclude this section with the remark that the simulation of
$\{S_{t}\}$ is straightforward. First, we generate a Brownian
trajectory, and form the associated Brownian bridge
$\{\beta_{tT}(\omega)\}$. We then select a value for $D_T$ by a
method consistent with the {\it a priori} probability density
$p(x)$, and substitute these in the formula $\xi_t(\omega)
=D_T(\omega)\int_0^t\sigma_s\rd s+ \beta_{tT} (\omega)$ for some
choice of $\{\sigma_t\}$. Finally, substitution of
$\{\xi_t(\omega)\}$ in (\ref{eq:7.1}) gives us a simulated path
$\{S_{t}(\omega)\}$. The statistics of the process $\{S_{t}\}$ are
obtained by repeating this procedure, the results of which can be
used to price derivatives, or to calibrate the information-flow rate
$\{\sigma_t\}$.

\section{Existence of the information process}
\label{sec:15}

We consider now what might be called the ``inverse problem'' for
information-based asset pricing. The idea is to begin with the
conditional density process $\{\pi_t(x)\}$ and to construct from it
the independent degrees of freedom represented by the $X$-factor
$D_T$ and the noise $\{\beta_{tT}\}$. The setup is as follows. On
the probability space $(\Omega, {\mathcal F},{\mathbb Q})$ let
$\{W_t\}$ be a Brownian motion and let $\{{\mathcal F}_t\}$ be the
filtration generated by $\{W_t\}$. Let $D_T$ be ${\mathcal
F}_T$-measurable, and let $\{\pi_t(x)\}$ denote the associated
conditional probability density process. We assume that
$\{\pi_t(x)\}$ satisfies the stochastic differential equation
\begin{eqnarray}
\rd\pi_t(x) = \nu_t(x-D_{tT})\pi_t(x)\, \rd W_t, \label{eq:6.1}
\end{eqnarray}
with the initial condition $\pi_0(x)=p(x)$, where $\{\nu_t\}$ is
given by (\ref{eq:3.4}), and where
\begin{eqnarray}
D_{tT}= \int_{0 }^\infty x\pi_t(x)\,\rd x.
\end{eqnarray}
We define the process $\{\xi_t\}$ as follows:
\begin{eqnarray}
\xi_t = (T-t)\int_0^t\frac{1}{T-s}\Big(\rd W_s + \nu_s D_{sT} \rd
s \Big). \label{eq:6.2}
\end{eqnarray}
Then we have the following result:

\vspace{0.4cm} \noindent {\bf Proposition~6}. {\it The random
variables $D_T$ and $\beta_{tT}=\xi_t- D_T\int_0^t \sigma_s\rd s$
are ${\mathbb Q}$-independent for all $t\in[0,T]$. Furthermore,
the process $\{\beta_{tT}\}$ is a ${\mathbb Q}$-Brownian bridge.}
\vspace{0.4cm}

Proof. To establish the independence of $D_T$
and $\beta_{tT}$ it suffices to verify that
\begin{eqnarray}
{\mathbb E}^{\mathbb Q}[{\rm e}^{x\beta_{tT}+yD_T}] = {\mathbb
E}^{\mathbb Q}[{\rm e}^{x\beta_{tT}}]\,{\mathbb E}^{\mathbb
Q}[{\rm e}^{yD_T}] \label{eq:6.3}
\end{eqnarray}
for arbitrary $x,y$. Using the tower property we have
\begin{eqnarray}
{\mathbb E}^{\mathbb Q}[{\re}^{x\beta_{tT}+yD_T}] = {\mathbb
E}^{\mathbb Q}\left[ {\re}^{x\xi_t} \,{\mathbb E}_t^{\mathbb
Q}\left[{\re}^{(y- x\int_0^t\sigma_s{\rm d} s) D_T}
\right]\right], \label{eq:6.4}
\end{eqnarray}
where we have inserted the definition of $\beta_{tT}$ given in the
statement of the Proposition. We consider the inner expectation
first. From equation (\ref{eq:3.6}) for the conditional
expectation of a function of $D_T$ we deduce that
\begin{eqnarray}
{\mathbb E}_t^{\mathbb Q}\left[ {\re}^{(y-x \int_0^t\sigma_s{\rm
d} s)D_T} \right] = \Phi_t^{-1} \int_{0}^\infty p(z)\,{\re}^{(y-x
\int_0^t \sigma_s{\rm d}s)z}\, \re^{ z\int_0^t \nu_u{\rm d}W_u^*-
\frac{1}{2} z^2 \int_0^t \nu_u^2{\rm d}u}\rd z, \label{eq:6.5}
\end{eqnarray}
where the process $\{\Phi_t\}$ is defined by (\ref{eq:7.4.1}). We
now change the probability measure from ${\mathbb Q}$ to ${\mathbb
B}$, so that the term $\Phi_t^{-1}$ appearing in (\ref{eq:6.5})
drops out to give us
\begin{eqnarray}
&& \hspace{-0.8cm} {\mathbb E}^{\mathbb Q}\left[ {\re}^{x\xi_t} \,
{\mathbb E}_t^{\mathbb Q}\left[{\re}^{(y- x\int_0^t\sigma_s{\rm d}
s)D_T}\right]\right] \nonumber \\ && \qquad = {\mathbb E}^{\mathbb
B} \left[\re^{x\xi_t}\int_0^\infty p(z)\,\re^{(y-
x\int_0^t\sigma_s{\rm d} s)z}\,\re^{z\int_0^t\nu_u{\rm d}W_u^*-
\frac{1}{2}z^2\int_0^t \nu_u^2 {\rm d}u } \rd z\right] \nonumber \\
&& \qquad = \int_{0}^\infty p(z)\,{\mathbb E}^{{\mathbb B}}\left[
\re^{x(T-t)\int_0^t\frac{1}{T-s}\, {\rm d}W_s^*+(y-x \int_0^t
\sigma_s{\rm d}s)z+z\int_0^t\nu_s {\rm d}W_s^*-\frac{1}{2}z^2
\int_0^t\nu_s^2{\rm d}s} \right]\rd z \nonumber \\ && \qquad =
\int_{0}^\infty p(z)\,\re^{(y-x\int_0^t\sigma_s{\rm d}s)z
-\frac{1}{2}z^2 \int_0^t\nu_s^2{\rm d}s+\frac{1}{2} \int_0^t
\alpha_s^2{\rm d}s}\,{\mathbb E}^{{\mathbb B}}\left[ \re^{\int_0^t
\alpha_s{\rm d}W_s^*-\frac{1}{2} \int_0^t \alpha_s^2{\rm d}s}
\right]\rd z, \label{eq:6.7}
\end{eqnarray}
where $\alpha_s=x(T-t)/(T-s)+z\nu_s$, and therefore
\begin{eqnarray}\label{eq:6.7a}
{\mathbb E}^{\mathbb Q}[{\rm e}^{x\beta_{tT}+yD_T}] =
\int_{0}^\infty p(z)\,\re^{(y-x\int_0^t\sigma_s{\rm
d}s)z-\frac{1}{2}z^2 \int_0^t\nu_s^2{\rm d}s+\frac{1}{2} \int_0^t
\alpha_s^2{\rm d}s}\rd z
\end{eqnarray}
Furthermore, making use of relation
(\ref{eq:3.8}) we have
\begin{eqnarray}
\exp\left(-xz\int_0^t\sigma_s{\rd}s -\half z^2 \int_0^t \nu_s^2
{\rd}s+\half \int_0^t \alpha_s^2{\rd}s\right) =
\exp\left(\frac{t(T-t)}{2T}\,x^2\right).
\end{eqnarray}
As a consequence, it follows from (\ref{eq:6.7a}) that
\begin{eqnarray}
{\mathbb E}^{\mathbb Q}\left[{\re}^{x\beta_{tT}+yD_T}\right] =
\left( \int_{0}^\infty p(z)\,\re^{yz}\rd z\right)\,
\exp\left(\frac{t(T-t)}{2T}x^2\right). \label{eq:6.8}
\end{eqnarray}
This establishes the independence of $\{\beta_{tT}\}$ and $D_T$.

The factorisation (\ref{eq:6.8}) also shows that the process
$\{\beta_{tT}\}$ is ${\mathbb Q}$-Gaussian, with mean zero and
variance $t(T-t)/T$. To establish that $\{\beta_{tT}\}$ is a
Brownian bridge, we must show that for $s\leq t$ the covariance of
$\beta_{sT}$ and $\beta_{tT}$ is given by $s(T-t)/T$. Alternatively,
it suffices to analyse the moment generating function ${\mathbb E}
[\re^{x\beta_{sT}+y \beta_{tT}}]$. We proceed as follows. First,
using the tower property we have
\begin{eqnarray}
{\mathbb E}^{\mathbb Q}\left[\re^{x\beta_{sT}+y\beta_{tT}}\right]
&=& {\mathbb E} \left[ \re^{x\xi_s+y\xi_t-(x\int_0^s\sigma_u{\rm
d}u+y\int_0^t \sigma_u{\rm d}u)D_T}\right]\nonumber \\ &=&
{\mathbb E} \left[ \re^{x\xi_s+y\xi_t}{\mathbb E}_t^{\mathbb
Q}\left[\re^{-(x\int_0^s\sigma_u {\rm d}u+y\int_0^t \sigma_u{\rm
d}u)D_T}\right] \right]. \label{eq:6.9}
\end{eqnarray}
Next, by use of formula (\ref{eq:3.6}), the inner expectation can
be carried out to give
\begin{eqnarray}
{\mathbb E}^{\mathbb Q}\left[\re^{x\beta_s+y\beta_{tT}}\right] =
{\mathbb E} \left[\re^{x\xi_s+y\xi_t} \Phi_t^{-1}\!\int_{0}^\infty
\!p(z)\, \re^{-(x\int_0^s\sigma_u {\rm d}u+y\int_0^t \sigma_u{\rm
d}u)z} \, \re^{ z \int_0^t \nu_u{\rm d}W_u^*- \frac{1}{2} z^2
\int_0^t \nu_u^2{\rm d}u}\rd z \right]. \label{eq:6.95}
\end{eqnarray}
If we change the probability measure to ${\mathbb B}$ the random
variable $\Phi_t$ in the denominator drops out, and we have
\begin{eqnarray}
{\mathbb E}^{\mathbb Q}\left[\re^{x\beta_s+y\beta_{tT}}\right] =
\int_{0}^\infty p(z) \, \re^{-(x\int_0^s \sigma_u{\rm
d}u+y\int_0^t \sigma_u{\rm d}u)z- \frac{1}{2} z^2 \int_0^t
\nu_u^2{\rm d}u}\, {\mathbb E}^{{\mathbb B}} \left[
\re^{x\xi_s+y\xi_t+z\int_0^t\nu_s{\rm d}W_s^*} \right]\rd z .
\label{eq:6.96}
\end{eqnarray}
Let us consider the inner expectation first. By defining
$a_u=x(T-s)/(T-u)$ and $b_u=y(T-t)/(T-u)+z\nu_u$ we can write
\begin{eqnarray}
{\mathbb E}^{{\mathbb B}}\left[ \re^{x\xi_s+y\xi_t+
z\int_0^t\nu_s{\rm d}W_s^*} \right] = {\mathbb E}^{{\mathbb
B}}\left[ \re^{\int_0^s a_u {\rm d}W_u^*+\int_0^t b_u{\rm
d}W_u^*}\right]. \label{eq:6.10}
\end{eqnarray}
However, since $\{W_t^*\}$ is a ${\mathbb B}$-Brownian motion,
using the properties of Gaussian random variable we find that
\begin{eqnarray}
{\mathbb E}^{{\mathbb B}}\left[ \re^{\int_0^s a_u {\rm
d}W_u^*+\int_0^t b_u{\rm d}W_u^*}\right] = \exp \left\{\half
\left(\int_0^s a_u^2 {\rm d}u + \int_0^t b_u^2 {\rm d}u+2\int_0^s
a_u b_u{\rm d}u\right) \right\}. \label{eq:6.105}
\end{eqnarray}
Substituting the definitions of $\{a_u\}$ and $\{b_u\}$ into the
right side of (\ref{eq:6.105}) and combining the result with the
remaining terms in the exponent of the right side of
(\ref{eq:6.96}) we find that the terms involving the integration
variable $z$ drop out, and we are left with the integral of the
density function $p(z)$, which is unity. Gathering the remaining
terms we obtain
\begin{eqnarray}
{\mathbb E}^{\mathbb
Q}\left[{\re}^{x\beta_{sT}+y\beta_{tT}}\right] = \exp \left\{\half
\left( x^2\,\frac{s(T-s)}{T} + y^2\,
\frac{t(T-t)}{T}+2xy\,\frac{s(T-t)}{T}
\right) \right\}.
\end{eqnarray}
It follows that the covariance of $\beta_{sT}$ and $\beta_{tT}$
for $s\leq t$ is given by
\begin{eqnarray}
\left.\frac{\partial^2}{\partial x\partial y}{\mathbb E}^{\mathbb
Q}\left[
{\re}^{x\beta_{sT}+y\beta_{tT}}\right]\right|_{x=y=0}=\frac{s(T-t)}{T}.
\end{eqnarray}
This establishes the assertion that $\{\beta_{tT}\}$ is a
${\mathbb Q}$-Brownian bridge. \hspace*{\fill} $\square$

The result above shows that, for the class of price processes we
are considering, even if at the outset we take the ``usual'' point
of view in financial modelling, and regard the price process of
the asset as being adapted to some ``prespecified'' filtration,
nevertheless it is possible to \emph{deduce} the structure of the
underlying information-based model.

\section{Multi-factor models with a time-dependent information
flow rate} \label{sec:16}

Let us now turn to consider the case of a single cash flow $D_T$
that depends on a \emph{multiplicity} of market factors
$\{X^{\alpha}_{T_k}\}^{\alpha=1,\ldots,m_k}_{k=1,\ldots,n}$, where
we have the $n$ pre-designated information dates
$\{T_k\}_{k=1,2,\ldots,n}$, and where for each value of $k$ we
have a set of $m_k$ market factors. For simplicity we set $T=T_n$.
Each market factor $X_{T_k}^{\alpha}$ is associated with an
information process
\begin{eqnarray}
\xi_{tT_k}^{\alpha} = X_{T_k}^{\alpha} \int_0^t
\sigma_{sT_k}^{\alpha} \rd s + \beta_{tT_k}^{\alpha},
\label{eq:2.25}
\end{eqnarray}
where $X_{T_k}^{\alpha}$ and $\{\beta_{tT_k}^{\alpha}\}$ are
independent. It should be evident that although the random
variable $D_T$ representing the cash flow is ${\mathcal
F}_T$-measurable, the values of some of the $X$-factors upon which
it depends may be revealed at earlier times. That is to say, the
uncertainties arising from some of the economic elements affecting
the value of the cash flow at time $T$ may be resolved before that
time.

Since the $X$-factors are independent, it follows that for each
market factor the associated conditional density process
$\pi_{tT_k}^{\alpha}(x)$ takes the form given in (\ref{eq:3.13}),
and the corresponding dynamical equation is given by
\begin{eqnarray}
\rd\pi_{tT_k}^{\alpha}=
\nu_{tT_k}^{\alpha}\left(x_k^{\alpha}-{\mathbb E}^{{\mathbb
Q}}\left[ X^{\alpha}_{T_k}\left|{\mathcal F}_t\right.\right]
\right)\,\pi_{tT_k}^{\alpha}\, \rd W_t^{\alpha k}. \label{eq:8.2}
\end{eqnarray}
The function $\nu^{\alpha}_{tT_k}$ appearing here is given by an
expression of the form (\ref{eq:3.4}):
\begin{eqnarray}
\nu^{\alpha}_{tT_k}=\sigma^{\alpha}_{tT_k}+\frac{1}{T_k-t}
\int^t_0\sigma^{\alpha}_{sT_k}\rd s.
\end{eqnarray}
The innovation process $\{W_t^{\alpha k}\}$ is defined in terms of
$\{\xi_{tT_k}^{\alpha}\}$ via a relation of the form
\begin{eqnarray}
W^{\alpha k}_t=\xi_{tT_k}^{\alpha}+\int^t_0\frac{1}{T_k-s}\,
\xi_{sT_k}^{\alpha}\rd s-\int^t_0\nu_{sT_k}^{\alpha}
X^{\alpha}_{T_k}\rd s.
\end{eqnarray}
The conditional expectation ${\mathbb E}^{\mathbb Q}[D_T|
{\mathcal F}_t]$ is thus given by the multi-dimensional integral
\begin{eqnarray}
D_{tT} &=& \int_{0}^{\infty}\!\!\cdots\!\int_{0}^{\infty}\!
\Delta_T(x^1_1,\ldots,x^{m_1}_1,\ldots,x^1_n,\ldots,x^{m_n}_n)
\nonumber\\ && \times \pi_{t1}(x_1^1)\cdots\pi_{t1} (x_1^{m_1})
\cdots \pi_{tn}(x_n^1)\cdots\pi_{tn}(x_n^{m_n})\,\rd x_1^1
\cdots\rd x_1^{m_1}\cdots\rd x_n^1\cdots\rd x_n^{m_n},
\end{eqnarray}
and the price of the asset for $t<T$ is $S_t=P_{tT}D_{tT}$. A
straightforward application of Ito's rule then establishes the
following result:

\vspace{0.4cm} \noindent {\bf Proposition~7}. {\it The price
process $\{S_t\}_{0\leq t<T}$ of an asset that pays a single
dividend $D_T$ at time $T(=T_n)$ depending on the market factors $
\{X_{T_k}^{\alpha}\}_{k=1,2,...n}^{\alpha=1,2,...,m_k}$, satisfies
the dynamical equation
\begin{eqnarray}
\rd S_{t} = r_t S_{t}\rd t + \sum_{k=1}^n\sum_{\alpha=1}^{m_k}
\nu_{tT_k}^{\alpha} {\rm Cov}_t [D_T,X_{T_k}^{\alpha}]\, \rd
W_t^{\alpha k}, \label{eq:8.4}
\end{eqnarray}
where
\begin{eqnarray}
D_T=\Delta_T\left(X^{\alpha}_{T_1},\ldots,
X^{\alpha}_{T_k}\right).
\end{eqnarray}
Here ${\rm Cov}_t[D_T,X_{T_k}^{\alpha}]$ denotes the covariance
between the cash-flow $D_T$ and the market factor
$X_{T_k}^{\alpha}$, conditional on the information ${\mathcal F}_t$
generated by the information processes $\{\xi_{sT_k}^{\alpha}
\}_{0\leq s\leq t}$}. \vspace{0.4cm}

In the more general case of an asset that pays multiple dividends
(see Section~\ref{sec:6}) the price is given by
\begin{eqnarray}
S_t=\sum^n_{k=1}{\mathds 1}_{\{t<T_k\}}P_{tT_k}{\mathbb
E}^{\mathbb Q} \left[ \left. \Delta_{T_k}
\left(\{X^{\alpha}_{T_j}\}_{j=1,\ldots,k}^{\alpha
=1,2,...,m_j}\right) \right|{\mathcal F}_t\right].
\end{eqnarray}

\vspace{0.4cm} \noindent {\bf Proposition~8}. {\it The price process
$\{S_t\}$ of an asset that pays the random dividends $D_{T_k}$ on
the dates $T_k$ $(k=1,\ldots,n)$ satisfies the dynamical equation
\begin{eqnarray}
\rd S_{t} = r_t S_{t}\rd t + \sum_{k=1}^n\sum_{\alpha=1}^{m_k}
{\mathds 1}_{\{t<T_k\}}\nu_{tT_k}^{\alpha} {\rm Cov}_t
[D_{T_k},X_{T_k}^{\alpha}]\,\rd W_t^{\alpha k} + \Delta_{T_k} \rd
{\mathds 1}_{\{t<T\}}, \label{eq:8.4a}
\end{eqnarray}
where
\begin{eqnarray}
D_{T_k}=\Delta_{T_k}\left(\{X^{\alpha}_{T_j}
\}^{\alpha=1,\ldots,m_j}_{j=1,\ldots,k}\right).
\end{eqnarray}
Here ${\rm Cov}_t[D_{T_k},X_{T_k}^{\alpha}]$ denotes the
covariance between the dividend $D_{T_k}$ and the market factor
$X_{T_k}^{\alpha}$, conditional on the market information
${\mathcal F}_t$.} \vspace{0.15cm}

We conclude that the multi-factor, multi-dividend situation is fully
tractable when the information-flow rates associated with the
various market factors are time dependent. A straightforward
extension of Proposition~8 allows us to formulate the joint price
dynamics of a system of assets, the associated dividend flows of
which may depend on common market factors. As a consequence, a
specific model for stochastic volatility and correlation emerges for
such a system of assets, and it is one of the main conclusions of
this paper that such a model can be formulated. The
information-based ``$X$-factor'' approach presented here thus offers
a new insights into the nature of volatility and correlation, and as
such may find applications in a number of different areas of
financial risk analysis. We have in mind, in particular,
applications to equity portfolios, credit portfolios, and insurance,
all of which exhibit intertemporal market correlation effects. We
also have in mind the problem of firm-wide risk management and
optimal capital allocation for banking institutions.

\vskip 15pt \noindent {\bf Acknowledgements}. The authors thank
T.~Bielecki, T.~Bj\"ork, I.~Buckley, H.~B\"uhlmann, S.~Carter,
I.~Constantinou, M.~Davis, J.~Dear, A.~Elizalde, B.~Flesaker,
H.~Geman, V.~Henderson, D.~Hobson, T.~Hurd, M.~Jeanblanc, A.~Lokka,
J.~Mao, B.~Meister, M.~Monoyios, M.~Pistorius, M.~Rutkowski,
D.~Taylor, and M.~Zervos for stimulating discussions. The authors
are also grateful for helpful comments made by seminar participants
at various meetings where parts of this work have been presented,
including: the Developments in Quantitative Finance conference, July
2005, Isaac Newton Institute, Cambridge; the Mathematics in Finance
conference, August 2005, Kruger National Park, RSA; the School of
Computational and Applied Mathematics, University of the
Witwatersrand, RSA, August 2005; CEMFI (Centro de Estudios
Monetarios y Financieros), Madrid, October 2005; the Department of
Actuarial Mathematics and Statistics, Heriot-Watt University,
December 2005; the Department of Mathematics, King's College London,
December 2005; the Bank of Japan, Tokyo, December 2005; and Nomura
Securities, Tokyo, December 2005. DCB acknowledges support from The
Royal Society; LPH and AM acknowledge support from EPSRC (grant
number GR/S22998/01); AM thanks the Public Education Authority of
the Canton of Bern, Switzerland, and the UK Universities ORS scheme,
for support.

\vskip 15pt \noindent {\bf References}.

\end{document}